\documentclass[a4paper,fleqn,usenatbib]{mnras}

\usepackage{savesym}
\usepackage{amsmath}
\savesymbol{iint}
\usepackage{txfonts}
\restoresymbol{TXF}{iint}

\usepackage[T1]{fontenc}
\usepackage{ae,aecompl}

\usepackage{graphicx}	
\usepackage{amssymb}	

\newcommand{\unit}[1]{\ifmmode\,{\rm #1}\else$\,{\rm #1}$\fi}

\newcommand{\etal}{~{et~al.}\ }  

\newcommand{\kms}{km~s$^{-1}$}
\newcommand{\msun}{$M_\odot$}
\newcommand{\mhi}{$M_{\rm \footnotesize\textsc{H\,i}}$}

\newcommand{\hi}{H{\sc\,i}}

\title[Extended \hi\ Features in the Leo Cloud]{ALFALFA and WSRT Imaging of Extended \hi\ Features in the Leo Cloud of Galaxies}
\author[Leisman\etal]{Lukas Leisman,$^{1}$
Martha P. Haynes,$^{1}$
Riccardo Giovanelli,$^{1}$
Gyula J\'{o}zsa,$^{2,3,4}$
\newauthor
Elizabeth A. K. Adams,$^{5}$
Kelley M. Hess$^{5,6}$
\\
$^{1}$Cornell Center for Astrophysics and Planetary Science, Space Sciences Building, Cornell University, Ithaca, NY 14853, USA\\
$^{2}$SKA South Africa Radio Astronomy Research Group, 3rd Floor, The Park, Park Road, Pinelands 7405, South Africa\\
$^{3}$Rhodes Centre for Radio Astronomy Techniques \& Technologies, Department of Physics and Electronics, Rhodes University,\\ PO Box 94, Grahamstown 6140, South Africa\\
$^{4}$ ArgelanderInstitut f\"ur Astronomie, Auf dem H\"ugel 71, D-53121 Bonn, Germany\\
$^{5}$ASTRON, the Netherlands Institute for Radio Astronomy, Postbus 2, 7990 AA, Dwingeloo, The Netherlands\\
$^{6}$Kapteyn Astronomical Institute, University of Groningen, 
PO Box 800, NL-9700 AV Groningen, The Netherlands
}

\date{Accepted 2016 August 17. Received 2016 August 12; in original form 2016 May 18}

\pubyear{2016}

\begin{document}
\label{firstpage}
\pagerange{\pageref{firstpage}--\pageref{lastpage}}
\maketitle

\begin{abstract}
We present ALFALFA \hi\ observations of a well studied region of the Leo Cloud, which includes
the  NGC~3227 group and the  NGC~3190 group.
We detect optically dark \hi\ tails and plumes with extents potentially exceeding 600~kpc, well beyond the field of view of previous observations. These \hi\ features contain $\sim$40\% of the total \hi\ mass in the  NGC~3227 group
 and $\sim$10\% in the  NGC~3190 group. We also present WSRT maps which show the complex 
morphology of the extended emission in the NGC~3227 group.
We comment on previously proposed models of the interactions in these groups, and
the implications for the scale of group processing through interactions.
Motivated by the extent of the \hi\ plumes, we place the \hi\ observations in the context of the larger loose group, 
demonstrating the need for future sensitive, wide field \hi\ surveys to understand the role of group processing
in galaxy evolution.

\end{abstract}


\section{Introduction}
\label{intro}

Most galaxies at z$\sim$0 can be found in group environments (e.g. \citealp{tago08a}), where evidence of tidal interactions and gas stripping are particularly prevalent (e.g. \citealp{hibbard01b}; \citealp{hibbard01a}). These observations of ongoing tidal interactions suggest that galaxies can undergo significant morphological evolution in the group environment, possibly playing a major role in the morphology-density relation (e.g. 
\citealp{postman84a}; 
\citealp{bekki11a}; \citealp{hess13a}). 

\hi\ observations can give direct evidence to the fundamental role of group processing in galaxy evolution, tracing the recent interaction history of group galaxies (e.g. \citealp{yun94a}). However, while the inner interactions of merging galaxies have been observed in detail, (e.g. \citealp{rand94a}), 
\hi\ studies covering the full extent of interactions in groups are difficult to execute due to the need for high sensitivity over wide fields; wide field \hi\ surveys have been limited in resolution and sensitivity, and interferometric studies have been limited to the field of their primary beam.

Still, \hi\ mapping on the group scale is important for understanding the ubiquity of these interactions, the fraction of gas involved, and the interaction time scales (e.g., \citealp{serra15a}), which in turn constrain the impact of tidal processing in groups on galactic evolution, relative to other processes, like the decrease in cold gas accretion \citep{sancisi08a}. 

The Arecibo Legacy Fast ALFA (Arecibo L-band Feed Array) survey (ALFALFA; \citealp{giovanelli05a}) provides high sensitivity, unbiased, wide field maps of \hi\ in the local volume, and has traced the atomic gas distribution in a variety of local groups (
 \citealp{stierwalt09a}; \citealp{lee-waddell12a}; \citealp{lee-waddell14a}). 

Here we present the ALFALFA data on a well studied loose group in the Leo Cloud, known to show significant evidence of interaction. The ALFALFA data are not limited to an arbitrary region on the sky, allowing us to search for additional structures and constrain the cold gas fraction outside stellar disks over the full $\sim$9~Mpc$^2$ physical region. These data reveal substantial intragroup cold gas, demonstrating the power of sensitive, wide field \hi\ mapping. 

We summarize prior work on this group in section \ref{GH58} and present our observations in section \ref{observations}. We present our results in section \ref{results}, discuss the group context in section \ref{discussion} and conclude in section \ref{conclusions}. For all calculations, the assumed cosmology is $H_0 = 70~\unit{km s^{-1} Mpc^{-1}}$,
$\Omega_m=0.3$, and $\Omega_\Lambda=0.7$.

\section{Placement in the Cosmic Web}
\label{GH58}

\begin{figure*}
\centering
\includegraphics[width=\textwidth]{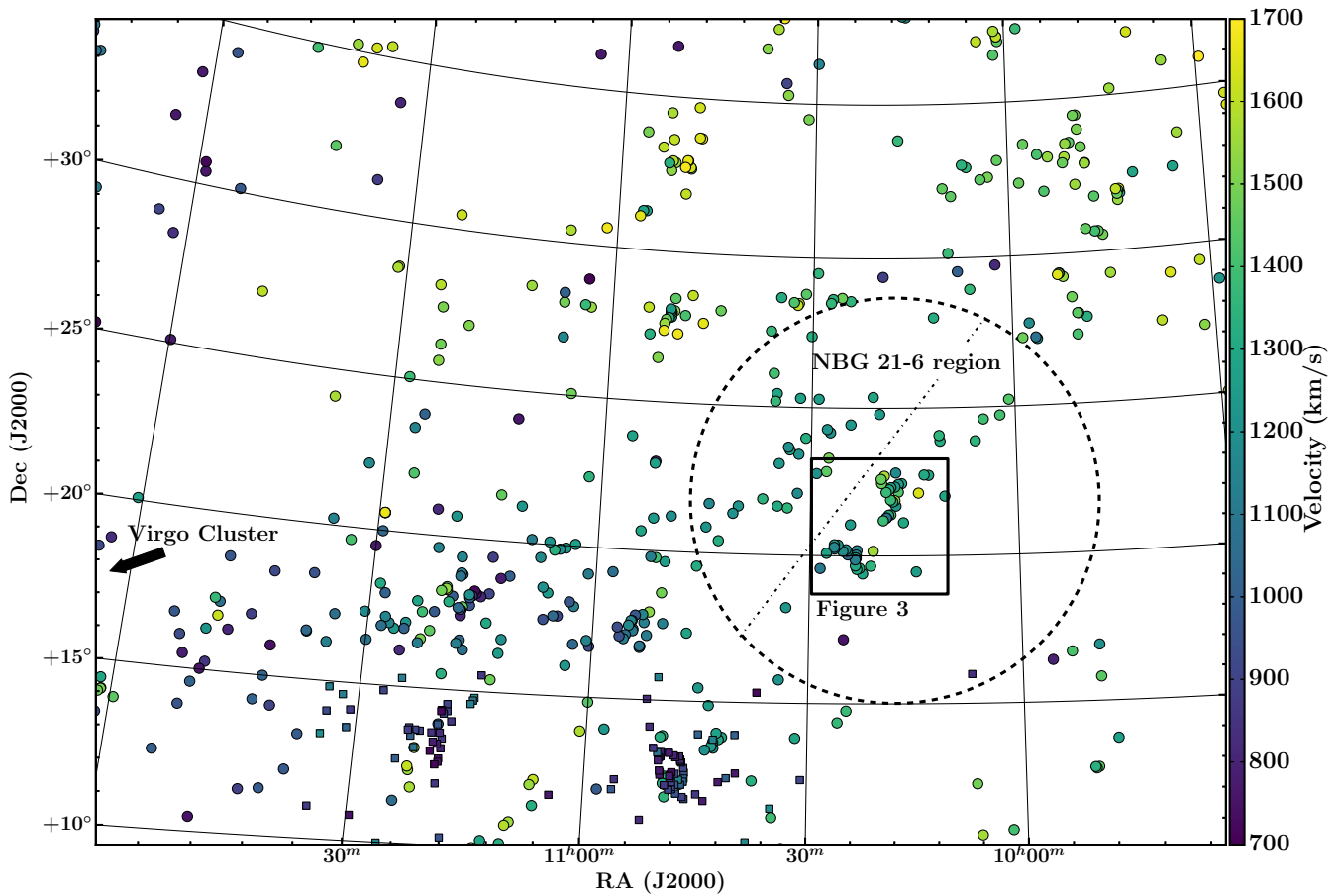}
\caption{
The distribution of galaxies in the NBG~21-6 region within the context of the Leo Cloud, a sparsely populated filament at D$\sim$20-30~Mpc stretching from the lower left of the image (near the Virgo Cluster) to the upper right. Sources are colour coded by heliocentric recessional velocity, and show some gradient from high to low velocities as the filament approaches the Virgo Cluster.   The dashed circle has a radius of 3~Mpc at a distance of 25~Mpc, and shows the location of the NBG 21-6 region that is the primary focus of this study. Some group catalogues divide the sources in the NBG 21-6 region into two subgroups, as indicated by the dashed-dotted line. The central/SW region (below the dashed-dotted line) itself has two main concentrations, the NGC~3227 and NGC~3190 groups (Figure \ref{bigmap}), which contain the large \hi\ structures discussed in section \ref{results}. Members of the foreground Leo Spur (D$\sim$10~Mpc), as identified by \citet{stierwalt09a}, are indicated by squares rather than circles.
}
\label{region}
\end{figure*}

\subsection{The NBG 21-6 Region} 
\begin{figure}
\centering
\includegraphics[width=7.5cm]{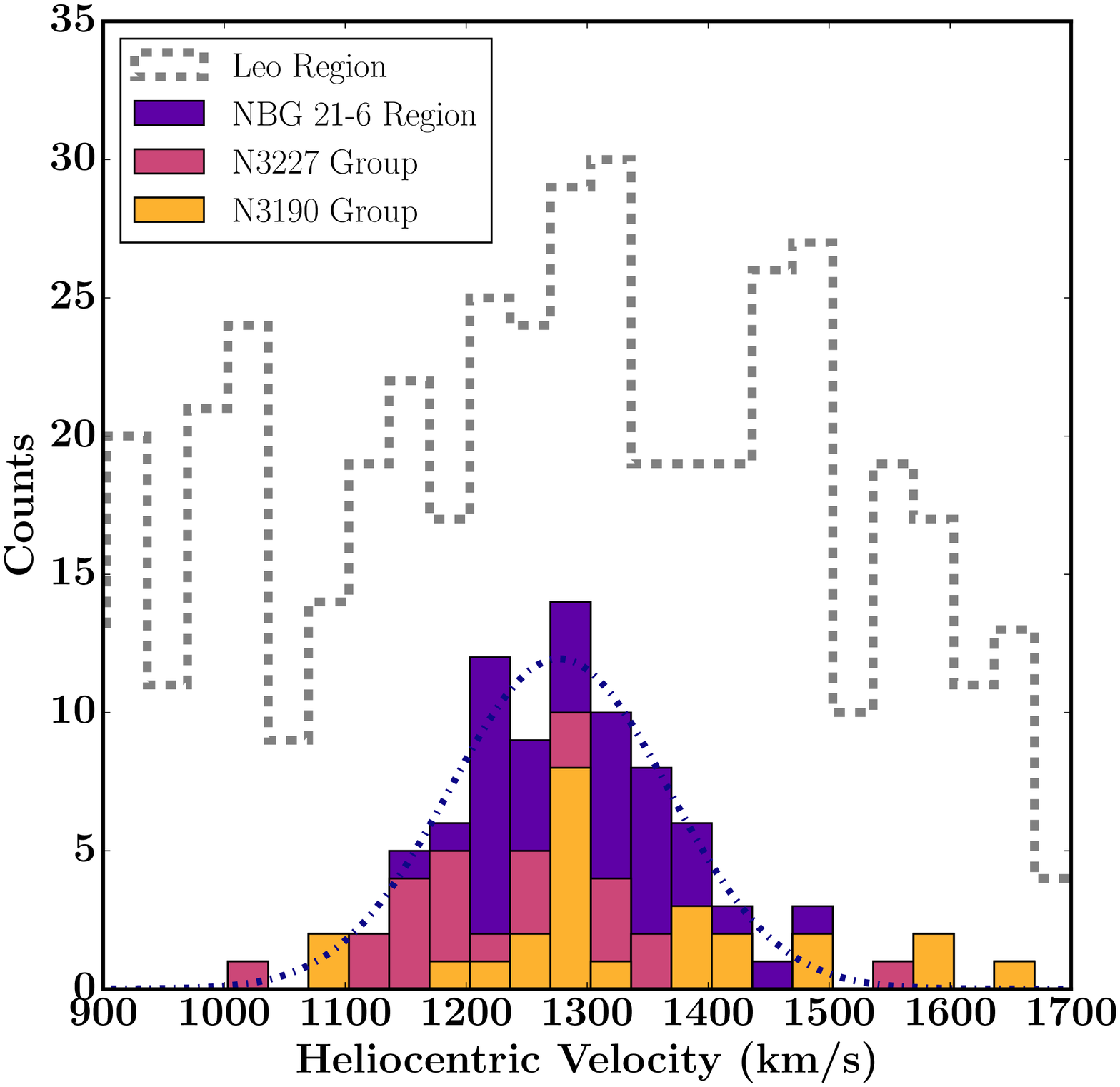} \\
\caption{Histogram of heliocentric recessional velocities in the Leo Region. The filled purple bars indicate the counts for the NBG 21-6 region (dashed circle in Figure \ref{region}), which shows a roughly Gaussian distribution (the best fitting Gaussian with $\sigma\sim$115~\kms\ is overplotted as a dash-dotted line), in contrast with the source distribution for all sources in Figure \ref{region}, indicated here by the dashed grey line. Sources in the  NGC~3190 (yellow) and  NGC~3227 (pink) groups are subsets of the NBG~21-6 distribution. 
}
\label{regionhist}
\end{figure}

Figure \ref{region}\footnotemark[1] shows the distribution of galaxies in the constellation Leo, between heliocentric recessional velocities of v$_h$=700~\kms\ and 1700~\kms. The primary structure in this velocity range is the Leo Cloud,\footnotemark[2] a loose collection of groups stretching $\sim$50$^{\circ}$ across the plane of the sky at a distance of $\sim$20-30~Mpc, from the edge of the Virgo Cluster around 12$^h$+15$^{\circ}$ (the lower left corner of Figure \ref{region}), to around 9$^h$30$^m$+35$^{\circ}$ (the upper right corner). A region of particular interest for \hi\ studies (see section \ref{GH58.HI}) is a loose association of galaxies spanning $\sim$6~Mpc in the NW part of the Leo Cloud (indicated with a dashed circle in Figure \ref{region}). Figure \ref{regionhist} shows the velocity distribution of \hi\ sources in this region, which is approximately Gaussian, with an average heliocentric recessional velocity of 1304~\kms, and a dispersion of $\sigma\sim$115~\kms, in contrast with the more uniform distribution of the Leo Cloud at large. 

\footnotetext[1]{Data are taken from the Arecibo General Catalog (AGC),  a private database maintained over the years by MPH and RG; it contains all bright galaxies and ones of known redshift as available
in NED with cz $<$ 18000~\kms\ in the ALFALFA volume, additional unpublished \hi\ results as they are acquired, and bright galaxies in other regions of the sky.}
\footnotetext[2]{Analysis of the Leo Cloud is complicated by the superposition of the foreground Leo Spur at $\sim$10~Mpc, which is poorly discriminated in velocity space due to infall toward the Virgo Cluster (e.g., \citealp{tully87a}; \citealp{karachentsev15b}). 
ALFALFA results for the southern part of the Leo Cloud were presented in \cite{stierwalt09a}, but the coverage there only extended to $\delta< 16^{\circ}$.}

Yet, the group relationships of galaxies in this region is unclear. 
Several studies assign the galaxies in this region to a single group (e.g., \citealp{huchra82a}; \citealp{garcia93a}), which we will refer to as NBG~21-6 following the nomenclature of the Nearby Galaxies Catalog \citep{tully87a}. However, depending on the length scale chosen to link the galaxies, other authors have divided this region into two subgroups (e.g., \citealp{turner76a}; \citealp{geller83a}), as indicated by the dash-dotted line in Figure \ref{region}, and still others further divide the SW subgroup  
(pictured in Figure \ref{bigmap}) into the  NGC~3227 group, and the  NGC~3190 group (e.g., \citealp{makarov11a}). 

Figure \ref{regionhist} also shows the velocity distributions of  the component NGC~3190 and NGC~3227 groups, which have mean recessional velocities (dispersions) of 1220 (99) and 1353 (150)~\kms\ respectively, in contrast with the distribution for the full NBG~21-6 region.  Groups were assigned by selecting all sources within the boxed regions in Figure 3 (described in Section \ref{results}) within the stated velocity range. Tables \ref{N3227table} and \ref{N3190table} list the basic properties of the galaxies assigned to each group by this simplistic method (sources assigned to different groups by a more sophisticated algorithm \citep{makarov11a} are noted with a star).  Both the shape and narrowness of the velocity distribution
 suggest a relationship between the sources in the NBG~21-6, even while the precise nature of that relationship is uncertain.

\subsection{Distances in the NBG 21-6 Region}

Distance determinations to member galaxies compound the problem of group membership. Standard Hubble flow would give distances between 15 and 22~Mpc, but the entire region is falling into Virgo at $\sim$200~\kms\ \citep{karachentsev15a}, and individual galaxies are falling into the local filament and groups (see section \ref{discussion.bigpicture}). 
More, redshift-independent distance measurements for this group are both complicated and confusing. 
There are three sources with measured primary distances:  NGC~3226 (23.6~Mpc) and  NGC~3193 (34~Mpc) via surface brightness fluctuations \citep{tonry01a} 
and  NGC~3190, which contains two type Ia supernovae that give distance estimates ranging from 18 \citep{amanullah10a}
to 31~Mpc \citep{szabo03a}
depending on calibration. However, given the strong interactions in this region, the primary distances
to each of these objects may be strongly effected by systematic errors (see \citealp{serra13a} for
a brief discussion).
Other redshift-independent distances are little help, as estimates based on scaling relations are complicated by interactions; for example,  estimates of distances to the merging pair  NGC~3226/  NGC~3227 range from 14.5~Mpc \citep{yoshii14a} to 43.5~Mpc \citep{blakeslee01a}. 

Thus, observations that bridge the gap between individual galaxies and the group structure are important for disentangling the complex relationships in this region. In section \ref{discussion} we argue that the sources in this region are gravitationally interacting, and likely to be at a similar distance. Using the corrections given in the NASA Extragalactic Database, based on the local velocity field model from \cite{mould00a}, we estimate a distance of 24.3~Mpc from the average velocity of the NBG 21-6 group. We thus choose to follow \cite{serra13a} and assume a distance of 25~Mpc for all galaxies in the region for the remainder of this paper. We further defend this choice in section \ref{discussion.bigpicture}.

\subsection{Interactions in the NBG 21-6 Region}
\label{GH58.HI}

Previous studies have found significant evidence of interactions between group members in the NBG~21-6 region. 
The NGC~3190 group centers around Hickson Compact Group 44 (HCG~44), a compact group of 
four galaxies of similar optical brightness. 
Several studies have found that the galaxies in HCG~44 appear to be \hi\ deficient (see section \ref{discussion.compactgroups}), and  NGC~3187 and NGC~3190 show strong morphological evidence of tidal interactions. 
Moreover, \cite{serra13a} report the discovery of a large $\sim$300~kpc \hi\ tail extending to the north and west of the group. 

Similarly, 
\cite{mundell95a} found two tails stretching 7\arcmin\ north and 16\arcmin\ south (51 and 116~kpc at D=25~Mpc) of the interacting pair  NGC~3226 and  NGC~3227 (also called Arp~94) in deep C and D array VLA imaging.
Optical imaging shows a complex set of faint stellar filaments, arcs, and loops, and \cite{appleton14a} suggest a complex interaction history based on a plethora of multi-wavelength data.

While previous observations have been sufficient to reveal the complex nature of both the  NGC~3227 and  NGC~3190 groups, they have been limited by the field of view of their observations. Here we present the the first high sensitivity, complete \hi\ maps of then entire region, corroborating and extending these results (see sections \ref{results.arp94} and \ref{results.hcg44} for the  NGC~3227 and  NGC~3190 subgroups respectively), and painting a global picture of \hi\ in the region. 

\section{Observations and Data}
\label{observations}

\subsection{\hi\ Data from the ALFALFA Survey}

The ALFALFA observations and data processing are described in detail in previous papers (\citealp{giovanelli05a}; \citealp{saintonge07a}; \citealp{martin09a}; \citealp{haynes11a}). In brief, ALFALFA employs a two-pass, fixed azimuth drift scan strategy, with a bandwidth of 100~MHz and a spectral resolution of 24.4kHz (5.3~\kms\ at z=0) before Hanning smoothing. The data are bandpass subtracted, calibrated, and then flagged interactively for radio frequency interference (RFI). Once this ``level 1" processing is complete, the data are gridded into cubes 2.4$^{\circ}$ on a side and spanning the full spectral bandwidth from -2000 to 18,000~\kms\ (though the cubes are split into four subcubes of 1024 spectral channels each for easy processing). Each cube was flatfielded and rebaselined, and corrected for residual telescope pointing errors. Sources are extracted using the methods of \cite{saintonge07a}, and then each grid is examined by eye to improve on the automatic detection algorithm at lower signal to noise ratios (SNRs) and to identify optical counterparts in Sloan Digital Sky Survey (SDSS) and Digitized Sky Survey 2 (DSS2) images; final source parameters are measured and catalogued interactively. The public 70\% ALFALFA catalogue has $>$25,000 high SNR extragalactic detections \citep{jones16a}.

\begin{figure*}
\centering
\includegraphics[width=1.0\textwidth]{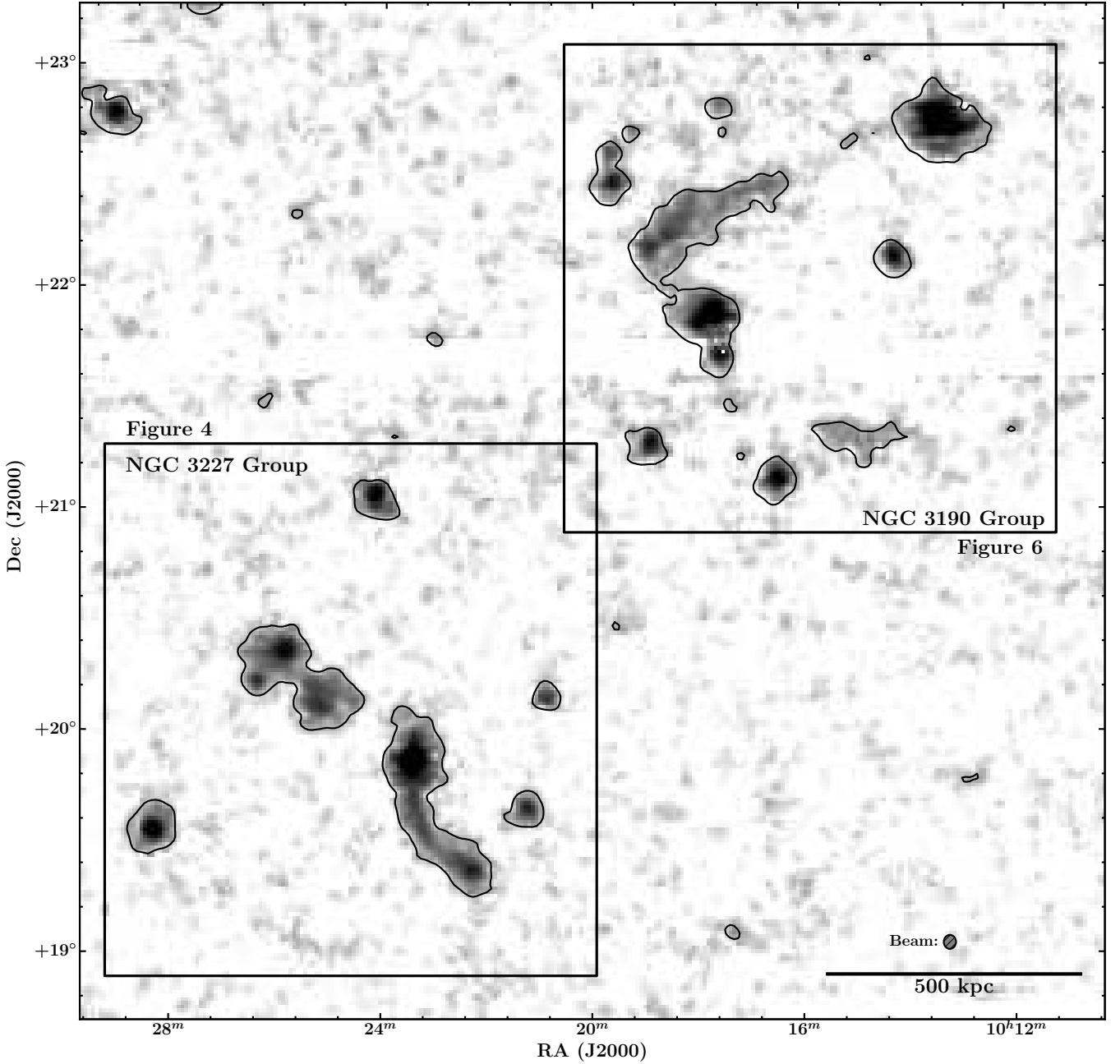}
\caption{%
$\sim$2x2~Mpc ALFALFA \hi\ moment 0 map of the region containing the  NGC~3190 and  NGC~ 3227 groups, from 911 to 1722\kms, with boxes indicating the approximate locations of the two groups, and black contours denoting emission 4$\sigma$ above the average rms noise of the image.
This map reveals that the \hi\ features in the groups approach the scale of the group. The extent of the features was previously undetected in WSRT and VLA synthesis observations because of their lower sensitivity, and restriction to a $\sim$0.5 degree primary beam. A few strips with no emission (especially around $\delta\sim$21$^d$+38$^m$) have been heavily masked due to significant RFI. 
}
\label{bigmap}
\end{figure*}

Due to their large angular extent, the structures discussed in this paper were originally split across multiple cubes. Thus, to better study the region on physically relevant scales, the ALFALFA data were regridded into a single 50 square degree cube centered at 10$^h$20$^m$ +21$^{\circ}$. 
The moment 0 map of the central portion of this region is shown in Figure \ref{bigmap}. This map was created by first smoothing to 2x the beam size and then masking pixels below 2.5$\sigma$.  Actual telescope scheduling, gain differences between the ALFA beams, and significant flagging due to RFI resulted in uneven integration times and rms values across the cube. To correct for this we used a weights map of effective integration time after flagging to create an rms cube which allowed us to determine the appropriate threshold for masking. One strip in particular, $\delta\sim$21$^d$38$^m$, was strongly affected by low level RFI and required additional masking. The mask was then applied to the unsmoothed cube, which was then summed over velocity channels from 911 to 1722~\kms, encompassing the full velocity range of the sources in the group. 

Total \hi\ fluxes for all sources in the region were extracted manually using the standard ALFALFA software (see \citealp{haynes11a}). 
However, the ALFALFA algorithm is optimized for point sources rather than highly extended and asymmetric sources like the features discussed below. Thus, 
all sources were remeasured by a modified version of the software which spatially integrates the spectrum over any arbitrary shaped area, and then divides by the summed value of the normalized beam over the same set of image pixels (e.g., \citealp{shostak80a}). This method gives consistent measurements with the standard software on point sources, and recovers up to $\sim$20\% more flux for the extended sources in this field.
 
\subsection{\hi\ Synthesis Imaging with WSRT}
Three sources in the vicinity of  NGC~3226/7 were included in an exploratory sample of synthesis observations studying extreme \hi\ sources without optical counterparts in ALFALFA (see \citealp{cannon15a} and \citealp{janowiecki15a} for more details). The connection to  NGC~3226/7 was noted, but the nature of the dark \hi\ knots, given their large separations from the merging pair, was unclear. 

We observed these three sources with 4x12h pointings with WSRT, 2 centered at 10$^h$25$^m$21.0$^s$ +20$^{\circ}$10\arcmin05\arcsec, one centered at 10$^h$22$^m$26.2$^s$ +19$^{\circ}$23\arcmin17\arcsec\ and one at 10$^h$25$^m$55.0$^s$ +20$^{\circ}$20\arcmin33\arcsec. The primary beams of the three pointing centers are 35\arcmin\ wide. 
We centered observations on the \hi\ line in one band of 10~MHz bandwidth and 1024 channels in two polarizations. This resulted in a broad range 
of line free channels for continuum subtraction and a velocity resolution of 4.1~\kms\ (FWHM) after Hanning smoothing.

We reduced the data using the same automated pipeline
as applied in \citet{wang13a}, originally used by \citet{serra12a}, as described in \citet{janowiecki15a}. 
The pipeline uses the data reduction software Miriad \citep{sault95a} wrapped into
a Python script. In brief, the pipeline automatically 
flags the data for RFI using a clipping method after filtering
the data in both the frequency and time domain. After primary
bandpass calibration, it iteratively deconvolves the data with the
CLEAN algorithm in order to apply a self-calibration, using CLEAN masks determined on the cube with
decreasing clip levels. We then subtract the continuum in the visibility domain and apply the  
calibration solution to the visibilities to then invert the data after Hanning smoothing. We then iteratively clean the data cubes down to the rms noise in the cubes, using CLEAN masks determined by filtering the data cubes with Gaussian kernels and
applying a clip level. We used a Robust weighting of r=0.4, and
binned the data to a velocity resolution of 12.3~\kms (FWHM; 6.0\kms\ channels) after Hanning smoothing.

For each cube we then created \hi\ total flux maps by smoothing the images to 2x the beam
size, masking any pixel below 3$\sigma$, applying the mask to the unsmoothed cubes, and then summing along the velocity axis.  
We calculate \hi\ column densities assuming optically thin \hi\ gas that fills the WSRT beam,
and also produce \hi\ moment 1 velocity maps from cubes masked at 3$\sigma$.
We measure the \hi\ flux by applying a mask based on the smoothed moment 0 map to cubes corrected for primary beam attenuation, and then extracting and fitting 1D spatially integrated \hi\ profiles. These fluxes are reported in Table \ref{N3227table}.  

\section{Results}
\label{results}

Figure \ref{bigmap} shows the ALFALFA moment 0 map of the central $\sim$20 deg$^2$ of the regridded cube. At 25~Mpc, this image covers $\sim$2x2~Mpc on the sky. 
The NGC~3227 and NGC~3190 groups are visible to the southeast and northwest of the center of the image, and are shown in greater detail in Figures \ref{arp94} and \ref{hcg44map} respectively. The \hi\ plumes associated with  NGC~3227 and the large feature in the  NGC~3190 group extend to similar scales as the groups themselves. Indeed, \cite{makarov11a} estimate the harmonic radii of the NGC~3227 and NGC~3190 groups to be 125~kpc and 276~kpc respectively (at D=25~Mpc), significantly less than the projected length of the structures discussed below.
The large extent of these \hi\ features is the main result of this paper. Here we describe the properties and extent of the features in the NGC~3227 and NGC~3190 groups in sections \ref{results.arp94} and \ref{results.hcg44}, and then discuss the mass of cold intragroup gas in section \ref{results.mass}. 

\subsection{Extended Tails in the  NGC~3227 Group} 
\label{results.arp94}

\begin{figure*}
\centering
\includegraphics[width=1.0\textwidth]{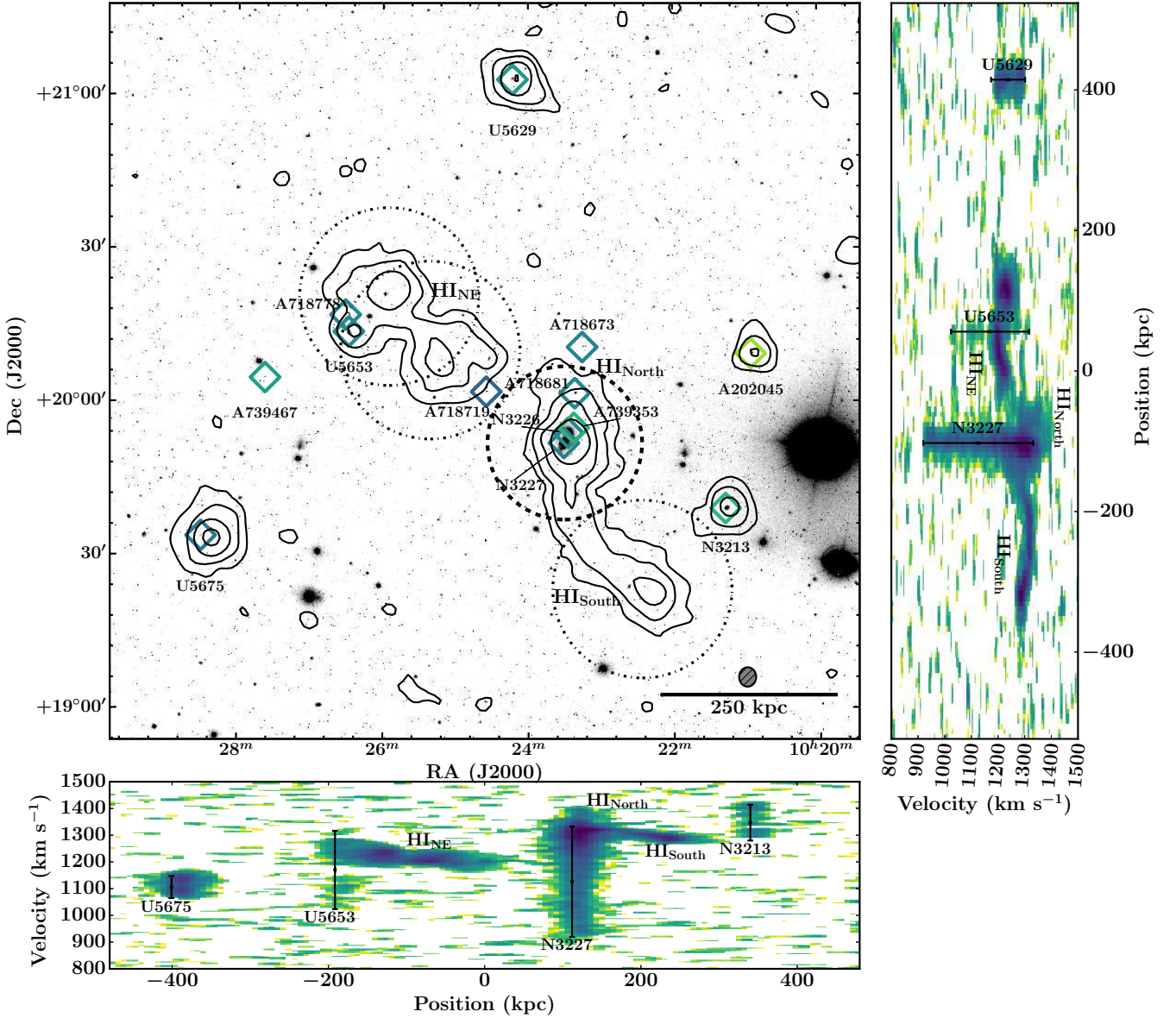}
\caption{ALFALFA moment 0 contours (summed over 900 to 1500~\kms) overlaid on a SDSS r-band optical image of the NGC~3227 group, demonstrating the large extent of the \hi\ features relative to the optical emission from the merging pair NGC~3226/7 (near the center of the image).  Contour levels are spaced logarithmically at 1.2, 2.4, 4.8, 9.6, and 19.2$\times10^{18}$ atoms~cm$^{-2}$ (assuming the \hi\ uniformly fills the ALFA beam of 3.3\arcmin x3.8\arcmin). The lowest contour level is 4$\sigma$ above the rms noise. The thick  dashed circle indicates the region mapped in higher resolution VLA imaging by \citet{mundell95a}, showing the need for wide field mapping for a full understanding of this system. The thinner dash dotted circles show the field of view of the WSRT observations shown in Figure \ref{Arp94WSRT}. Optical galaxies with known redshifts between 700 and 1700~\kms\ are marked with open diamonds, coloured by their recessional velocity to match the colour scale from Figure \ref{region}. Catalog designations AGC, UGC, and NGC are abbreviated by their first initial. 
The ALFA beam is represented by a hashed circle in the lower right. Right and Bottom: ALFALFA position-velocity diagram summing the \hi\ emission across the image from 10:27:30 to 10:21:45 in RA and from 19:00:00 to 20:40:00 in Dec respectively (the ranges are selected to minimize confusion and highlight the \hi\ plumes in the diagrams). Black vertical and horizontal bars indicate the ALFALFA velocity width at the 50\% flux level for detected galaxies in the region. The diagrams show the coherence of the plumes in velocity space, and the separation in velocity of HI$_{\rm NE}$ at $\sim$1200~\kms\ and HI$_{\rm North}$ at $\sim$1400~\kms .  
}
\label{arp94}
\end{figure*}

Figure \ref{arp94} shows the ALFALFA data for the NGC~3227 Group, overlaid on an SDSS r-band image created with Montage. The bottom and side panels show RA-velocity and Dec-velocity views of the group, with black vertical and horizontal bars indicate the ALFALFA velocity width at the 50\% flux level for detected galaxies. A dashed thick black circle indicates the region imaged by \cite{mundell95a}, who report the detection of two \hi\ plumes (labelled HI$_{\rm North}$ and HI$_{\rm South}$ in Figure \ref{arp94}) extending 51 and 116~kpc to the north and south of NGC~3227 in the center of the NGC~3227 group. However, the primary beam of the VLA at 1420~MHz (the effective field of view of the interferometer) is only $\sim$35\arcmin. Thus, while, the ALFALFA data confirm the plumes detected in the high resolution VLA maps, they find that the plumes extend far beyond the VLA primary beam to $\sim$10 and $\sim$40 arcminutes (73 and 291~kpc at 25~Mpc). They also reveal the presence of a previously unreported feature stretching 0.9
degrees to the NE (labelled HI$_{\rm NE}$ in Figure \ref{arp94}), distinct from the other plumes. 

All three \hi\ appendages appear to have no detectable stellar counterparts at the surface brightness limit of SDSS ($\sim$25 mags~asec$^{-2}$); 
optically detected galaxies of known redshift (i.e. of
m$_g <$17.7) are indicated with open diamonds, color coded by their velocity to match the scale in Figure \ref{region}. While there are several reasonably bright optical sources in this region, none of them are clearly associated with the \hi\ features.

%

\subsubsection{The Southern \hi\ Plume (HI$_{South}$)}
\begin{figure*}
\centering
{\bf a \hspace{3.3in}  b}\\
\includegraphics[width=7.5cm]{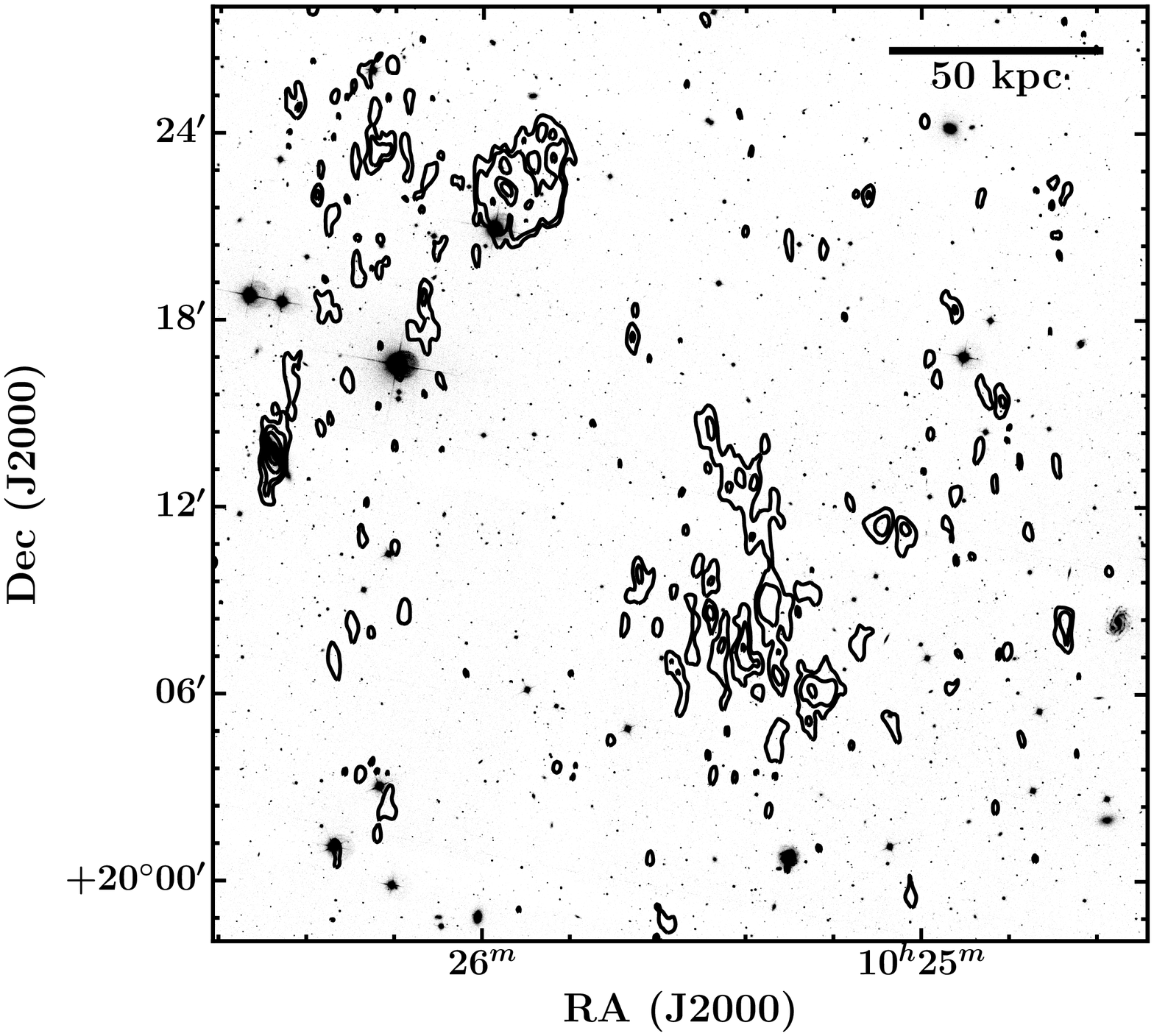}
\includegraphics[width=8.5cm]{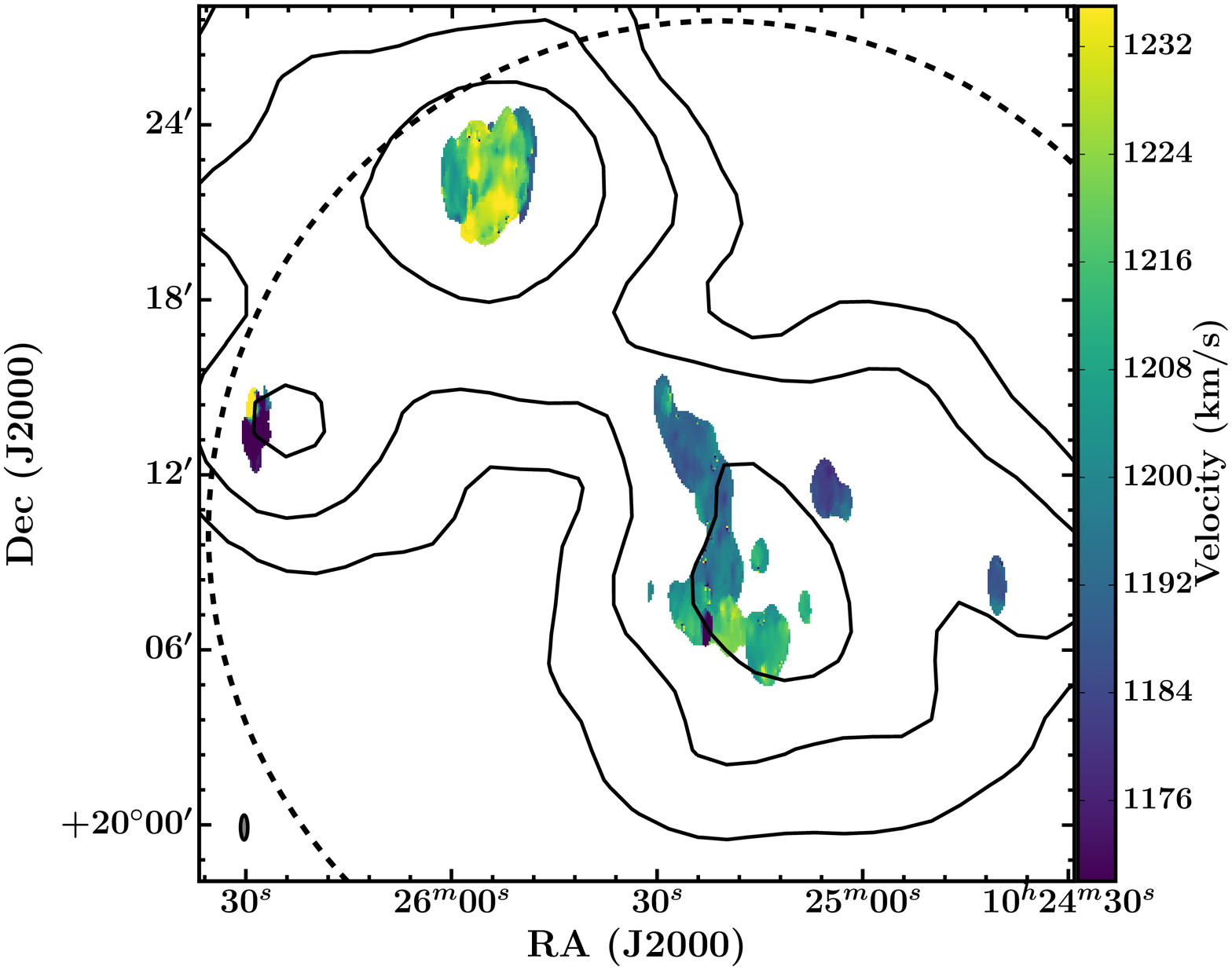} \\
{\bf c \hspace{3.3in}  d}\\
\includegraphics[width=7.5cm]{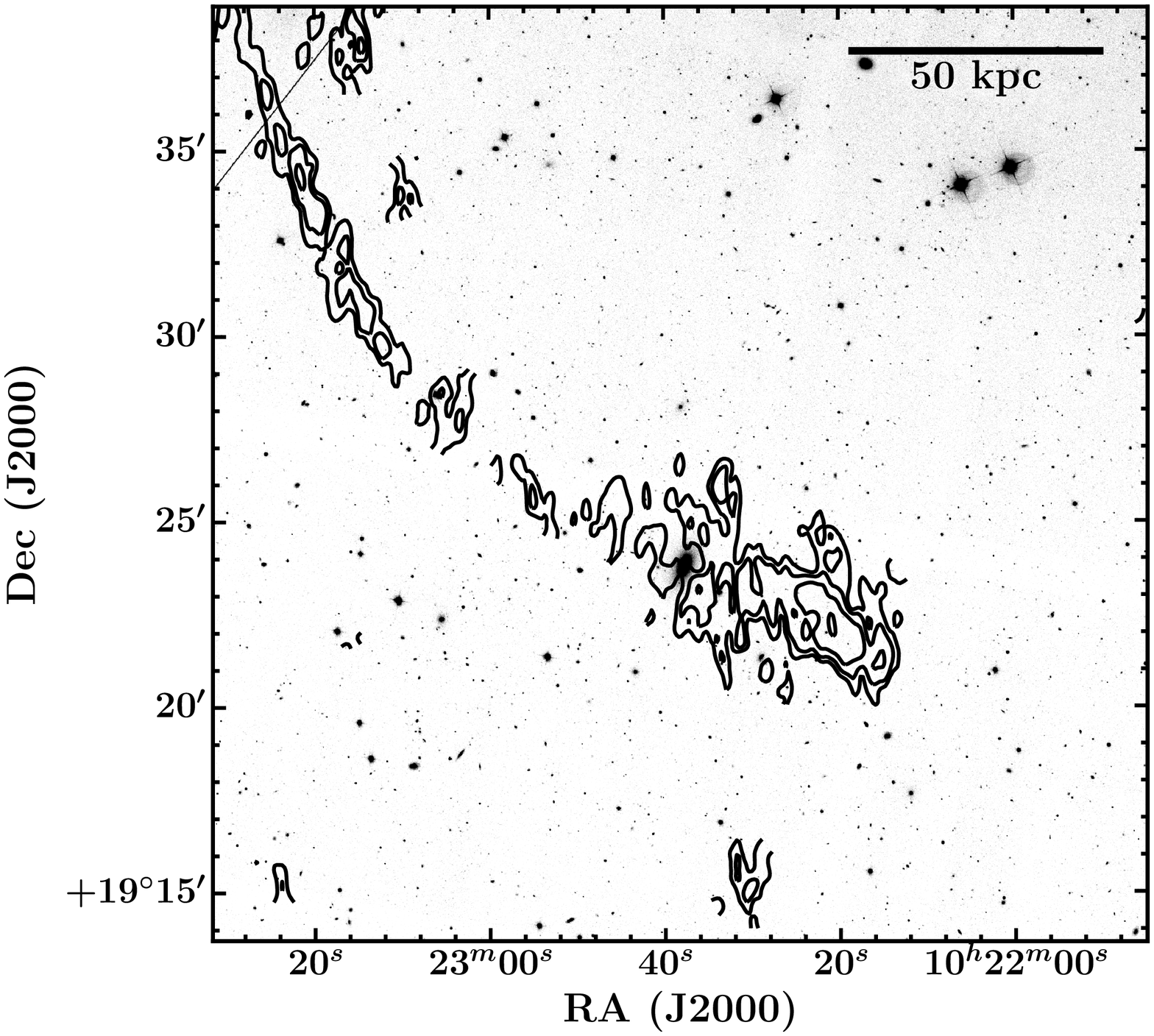}
\includegraphics[width=8.5cm]{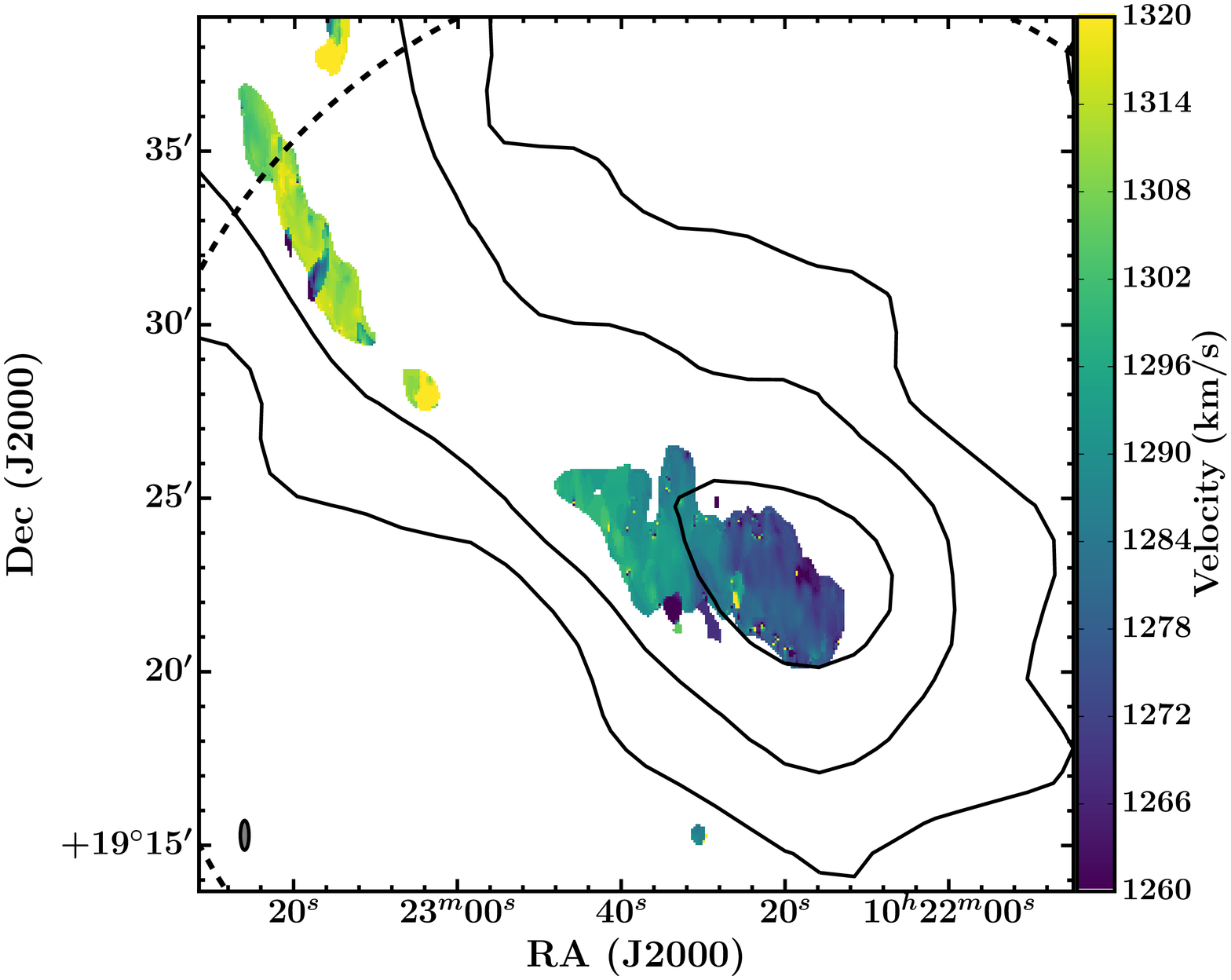}
\caption{Left: High resolution WSRT synthesis imaging of the ends of the Northeastern (HI$_{\rm NE}$) and Southern (HI$_{\rm South}$) \hi\ features associated with NGC~3227, showing messy, apparently tidal morphology in the higher column density gas. Panels a and c show WSRT \hi\ column density contours at 3.5, 7, 14, 28, and 56 $\times10^{19}$ atoms~cm$^{-2}$ (assuming the \hi\ uniformly fills the beam of 14\arcsec $\times$ 47\arcsec) of HI$_{\rm NE}$ and HI$_{\rm South}$ respectively, overlaid on g-band images from SDSS created with Montage. Panels b and d show WSRT moment 1 maps of HI$_{\rm NE}$ and HI$_{\rm South}$ respectively, with ALFALFA flux density contours from Figure \ref{arp94} overlaid in black. The WSRT observations match well with the lower resolution ALFALFA observations, and both features show coherent velocity structure. The FWHM of the WSRT primary beam is indicated by black dashed circles; significant primary beam corrections are necessary to estimate the flux of the entire features.  The WSRT beam is indicated by a small grey circle in the lower left corner of panels b and d.
}
\label{Arp94WSRT}
\end{figure*}

The \hi\ plume extending to the south of NGC~3227 shows excellent agreement with the \cite{mundell95a} maps within the region imaged by the VLA. 
Both data sets show that the plume (which we will refer to as HI$_{\rm South}$) rises steeply in velocity from $\sim$1200 \kms\ to $\sim$1300 \kms\ moving south away from NGC~3227. 
However, the feature extends $\sim$28\arcmin\ beyond the region covered by the VLA, continuing to rise
in velocity, and decrease in projected column density. $\sim$17\arcmin\ from NGC3226/7, it bends west and begins to decrease in velocity from $\sim$1320 to $\sim$1270~\kms. As shown in the PV panels in Figure \ref{arp94}, the end of the plume shows a marked increase in projected column density, and the velocity dispersion of the tail appears to decrease significantly moving south from  NGC~3227. However, the significantly higher resolution WSRT observations of the end of the plume, shown in panels c and d of Figure \ref{Arp94WSRT}, show a messy, extended distribution, inconsistent with any suggestion of a ``dark" or tidal dwarf galaxy (TDG). Indeed, the higher column density emission traces the ALFALFA data well, and has clear elongated, tail like morphology. The column density peaks at 2.3$\times10^{20}$ atoms cm$^{-2}$ (for a beam of 14\arcsec $\times$ 47\arcsec), but there is no associated stellar emission at the detection limits of SDSS. There is a low surface brightness galaxy at 10$^h$22$^m$53.2$^s$ +19$^{\circ}$34\arcmin36\arcsec\ without a measured redshift, but there is no associated \hi\ in the higher resolution WSRT images. The bright elliptical galaxy at 10$^h$22$^m$37.7$^s$ +19$^{\circ}$23\arcmin49\arcsec\ has a measured redshift of 11,792~\kms.

We note that NGC~3213 is $\sim$22\arcmin\ ($\sim$160 kpc) from the end of HI$_{\rm South}$, has a similar recessional velocity (see the RA-velocity plot in Figure \ref{arp94}), and 
appears to show some extended \hi\ emission at low SNR; however, our data do not have the sensitivity to determine if it is related to this southern plume.  
Thus, the \hi\ distribution mapped here is consistent with the properties of a tidal tail associated with the merging pair NGC~3226/7, but we cannot completely rule out the rather unlikely possibility that HI$_{\rm South}$ is instead an \hi\ bridge.

\subsubsection{The Northern \hi\ Plume (HI$_{North}$)}

The northern plume reported in \cite{mundell95a} also is consistent with the ALFALFA data, but is blended with the emission from the TDG identified in \cite{mundell04a} and  NGC~3227, in Figure \ref{arp94}. While it appears to spatially connect with the northeastern feature (HI$_{\rm NE}$, discussed below), the two tails
are well separated in velocity space, as demonstrated in the PV diagrams in Figure \ref{arp94}. The northern structure increases in velocity from 1300 to 1400~\kms, as one moves away from NGC~3226/7 to a projected linear separation of 15\arcmin\ (compared with HI$_{\rm NE}$ (section \ref{results.arp94.NE}), which approaches NGC~3226/7 at a recessional velocity of 1240~\kms).

 \cite{appleton14a} use Spitzer IRAC observations together with the VLA \hi\ data to connect this northern structure to the elliptical galaxy  NGC~3226, suggesting that the gas is infalling onto  NGC~3226. We note that it shows a significantly wider velocity dispersion ($\sim$50 \kms), and steeper velocity gradient than HI$_{\rm South}$, and that there are suggestions of significant substructure at lower SNR. There are two blue, fuzzy galaxies of moderate magnitude (AGC~718681 with M$_r$ = -17.3 and AGC~718673 with M$_r$ = -16.1 at 25~Mpc) that fall within this plume's \hi\ contours, but both objects
have recessional velocities that fall below that of the detected \hi\ (1151~\kms\ and 1166~\kms\ respectively). The compact dwarf AGC~739353 near NGC~3226 has a recessional velocity of 1338~\kms, somewhat higher than HI$_{\rm North}$. Further modelling will be necessary to confirm the origins of this feature (and if, in fact it is a tidal tail, a tidal bridge, or something else).

\subsubsection{The Northeastern Feature (HI$_{NE}$)}
\label{results.arp94.NE}
Figure \ref{arp94} also shows a bright \hi\ feature extending 0.9 degrees to the northeast of NGC~3226/7.
The feature is somewhat more massive than HI$_{\rm South}$, but has a clumpier gas distribution, higher velocity dispersion,
and more complicated velocity field. 
The nature of the feature's connection to NGC~3226/7 was not immediately clear, since it is not connected at the column density sensitivity of our data.  However, emission detected at 2.5$\sigma$ in at least 3 neighbouring beams leads away from NGC~3226/7 at around 1230~\kms\ (most easily seen in the PV diagrams), forming a smooth arc with the rest of the NE tail.  
This feature shows two primary projected density peaks in the ALFALFA data, one roughly associated with 
the velocity minimum, and the other near the end of the tail. The WSRT observations, shown in Figure \ref{Arp94WSRT}, resolve these peaks into several clumps, revealing complicated structure in the high column density emission.  The SW clump appears messy with two or more density peaks, and the velocity field suggests it may contain significant substructure. Some of the emission extends in the NE-SW direction, while the emission at higher velocity is elongated in the EW direction. The NE structure appears as a single clump with a peak column density of 3.9$\times10^{20}$~cm$^{-2}$ in the WSRT data (assuming the flux fills the 13\arcsec$\times$41\arcsec\ beam integrated over the width of the source), but does not appear to show coherent velocity structure.  

The disordered morphology at high column density, and the smooth, correlated nature of the structures in the ALFALFA data point to a tidal connection between NGC3226/7 and this NE structure. However, detailed modelling will be required to understand the full history of this feature, which may be the product of several interactions within the group. 

The entire structure shows no stellar emission in SDSS imaging.
There are 3 optical galaxies with measured redshifts in the nearby vicinity. AGC~718719 is
fairly small and blue, and appears in projection near NGC~3226/7. However, with a recessional velocity of 1028~\kms\ it appears to be unrelated
to the details of the tail. AGC~718778 (v$_{\rm helio}$=1163~\kms) is a low surface brightness dwarf located
near the northeastern end of the tail, and at a more similar velocity to the tail,
but is still at a lower recessional velocity than the detected gas. UGC~5653 is an edge on 
spiral with a clear dust lane located in the vicinity, but to the south of HI$_{\rm NE}$. 
While some of the emission from UGC~5653 is blended with HI$_{\rm NE}$, its central recessional velocity is lower than bulk of the nearby emission from HI$_{\rm NE}$ (1170~\kms\ when separated from the other emission). It is possible that HI$_{\rm NE}$ is, in fact, a bridge between UGC~5653 and NGC~3227, however the WSRT observations do not indicate a clear connection between the \hi\ in UGC~5653 and HI$_{\rm NE}$, and the geometry seems rather unfavourable. Further, without measured primary distances, velocity crowding from infall requires that these sources be interpreted with caution since there is a small chance that they are simply projected to a similar region of phase space.   However,
we do note that the 6 objects detected without any \hi\ in this subgroup (besides NGC~3226; listed in the bottom of Table \ref{N3227table}) are all
in close proximity to HI$_{\rm NE}$ or HI$_{\rm North}$ discussed above, 
and have somewhat similar optical colours and morphologies. 
All other objects brighter than m$_g$=17.7 have detected \hi\ in ALFALFA.

The moment 0 map also reveals suggestive tidal features at low SNR extending to
the southwest of the NE plume toward the low surface brightness dwarf galaxy UGC~5675. 
However, deeper observations would be necessary to put this low SNR suggestion on firmer footing. Understanding the origins of the three main features detected here will require detailed dynamical modelling.

\subsection{Extended Tails in the  NGC~3190 Group} 
\label{results.hcg44}

\begin{figure*}
\centering
\includegraphics[width=\textwidth]{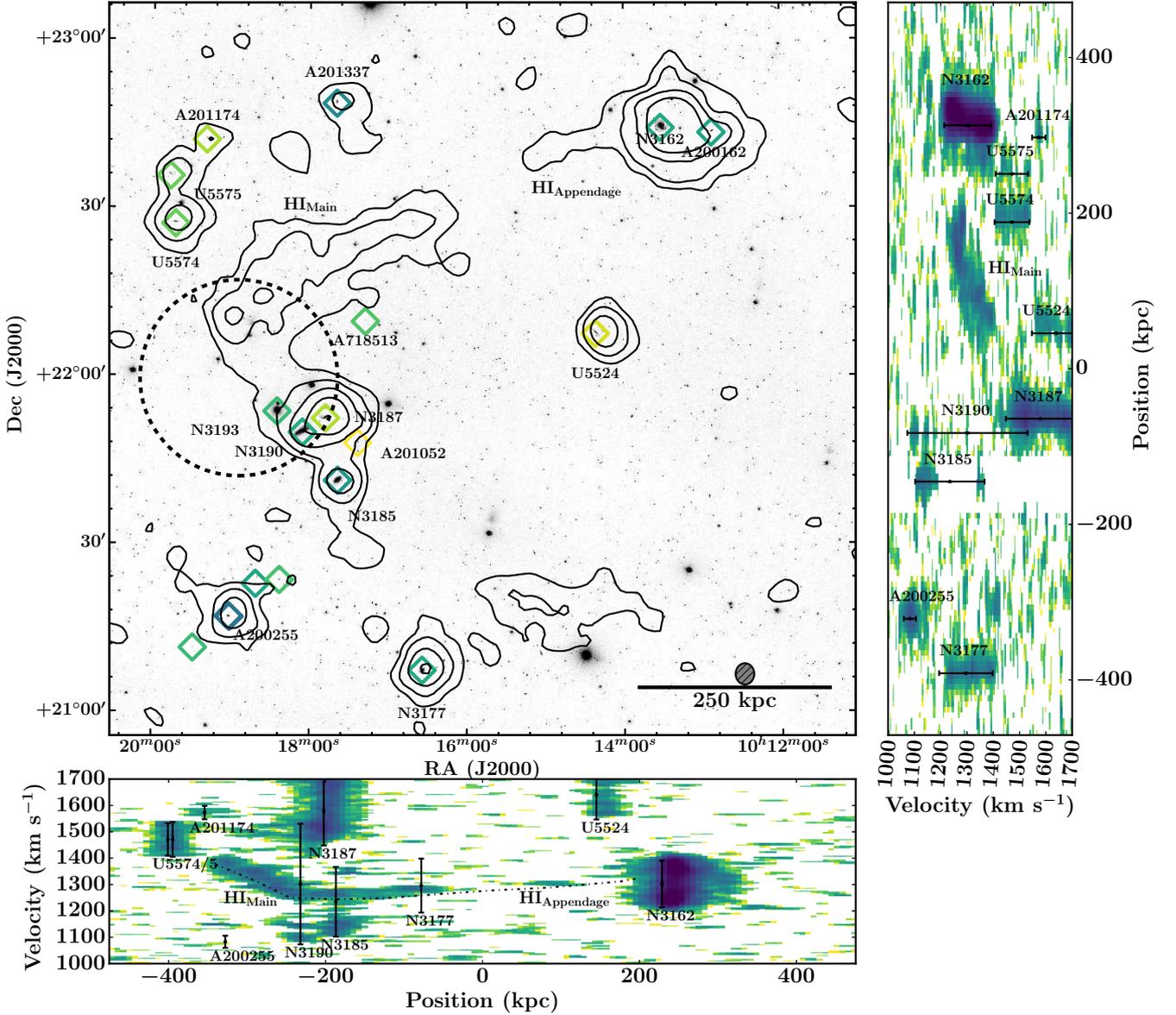}\\
\caption{ALFALFA imaging of the central region of the  NGC~3190 subgroup overlaid on an SDSS r-band optical image, showing the extended dark feature to the north of  HCG~ 44, and the likely connection to  NGC~ 3162. Optical galaxies with measured redshifts are indicated by open diamonds as in Figure \ref{arp94}. The ALFALFA moment 0 contour levels are spaced logarithmically at 2, 6, 18, and 54$\times10^{18}$ atoms~cm$^{-2}$ (assuming the \hi\ uniformly fills the ALFA beam of 3.3\arcmin$\times$3.8\arcmin). The lowest contour level is 4$\sigma$ above the rms noise. The black dashed circle indicates the FWHM of the WSRT primary beam from \citet{serra13a}. Note that the entire northern tail appears optically dark at the detection limits of SDSS. Right and Bottom: ALFALFA PV diagrams summing the \hi\ emission across the image from 10$^h$12$^m$ to 10$^h$20$^m$ in RA and from 21$^{\circ}$30\arcmin\ to 23$^\circ$ in Dec respectively (the ranges are selected to minimize confusion in the diagrams). Note that NGC~3185 is missing significant emission between v$\sim$1200-1300~\kms\ due to RFI masking, so the entirety of the main \hi\ tail is visible in the bottom PV diagram. The emission from HI$_{\rm Main}$ and HI$_{\rm Appendage}$ are underscored by a dashed dotted line, showing that they align well in both position and velocity. 
}
\label{hcg44map}
\end{figure*}

Located approximately 1~Mpc to the NW of the NGC~3227 group, the NGC~3190 group is the other primary overdensity in the NBG~21-6 region. Like the NGC~3227 group, the NGC~3190 group is know to exhibit extended \hi\ without apparent associated stellar emission. Specifically,  using a deep 6x12h pointing with WSRT,
 \cite{serra13a} reported the detection of a large \hi\ feature (which they refer to as the northern tail, T$_{\rm N}$) near the center of the group, extending $\sim$20\arcmin\ (220~kpc) to the northwest of  HCG~44, and optically dark down to $\mu_g$ = 28.5 mags~asec$^{-2}$ (in deep CFHT imaging). They further note the detection of HIPASS emission beyond the edge of the WSRT primary beam, which might extend the tail to $\sim$300~kpc, and the detection of a small cloud (C$_{\rm S}$) to the east of HCG~44 and the south of the main \hi\ feature. They suggest that the large \hi\ feature originated due to tidal stripping, and suggest that it may have originated from an interaction between NGC~3190 and NGC~3187, or from an interaction between NGC~3162 and the members of the compact group during a close flyby. 

Like in the NGC~3227 group, ALFALFA finds that this previously detected \hi\ feature extends to the group scale, well beyond the region previously studied, and finds evidence of additional extended \hi\ structures. 
Figure \ref{hcg44map} shows ALFALFA \hi\ column density 
contours overlaid on a Montage SDSS g-band mosaic of the NGC~3190 group.  The primary beam from \cite{serra13a} is shown as a black dashed circle. The ALFALFA observations, which have an effective integration time of 40 seconds per beam, are sensitive to low column density structure across the entire group, and show that full extent of the primary \hi\ structure. This feature, which we will refer to as HI$_{\rm Main}$ hereafter, contains all of T$_{\rm N}$ and the suggestive HIPASS emission, and more, covering $\sim$45\arcmin\ (330~kpc) across the sky. The feature shows significant substructure, which is confirmed in a recent, deep pointing with KAT-7, and discussed in detail in Hess et al., submitted. 

The ALFALFA data further show a low column density extension stretching SE from  NGC~3162 in the direction of HI$_{\rm Main}$. This appendage (HI$_{\rm Appendage}$) is relatively low signal to noise, but is detected at $>$2.5$\sigma$ in 6 contiguous beams, and has a narrow (W$_{50}$=31~\kms), but coherent structure in velocity space. Moreover, it is not simply a sidelobe of  NGC~3162, since it extends $\sim$30\arcmin\ from the center of  NGC~3162 (the peak of the first ALFA sidelobe is at $\sim$5\arcmin). The emission from the  NGC~3162 extension does not connect directly to HI$_{\rm Main}$ at the sensitivity of the ALFALFA data, however, Figure \ref{hcg44map} demonstrates that they are well aligned both spatially and in velocity space, strongly suggesting that HI$_{\rm Main}$ is, in fact connected to NGC~3162. The apparent connection of the \hi\ features means that the entire structure is nearly 1.4~degrees (610~kpc) long. 

Several other weaker pieces of evidence fit with the interpretation that HI$_{\rm Main}$ is associated with NGC~3162. The mass budget, presented below in section \ref{results.mass}, argues for the reality of the connection to  NGC~3162, since the tail contains 20\% of the \hi\ mass of  NGC~3162, but at least 40\% of the \hi\ mass of any of the sources in  HCG~44. Additionally,
NGC~3162 shows some sign of the presence of a forward tail encompassing the smaller spiral AGC~200162, and 
the gas in the full \hi\ structure appears to spread out spatially as it moves away from  NGC~3162, possibly consistent with gas spreading as a function of time after leaving  NGC~3162. 
However, we note that the velocity dispersion smoothly increases from $\sim$20~\kms\ to $\sim$60~\kms\ moving away from NGC~3162, which seems to suggest energy input from HCG~44. 

\begin{figure*}
\centering
\includegraphics[width=1.0\textwidth]{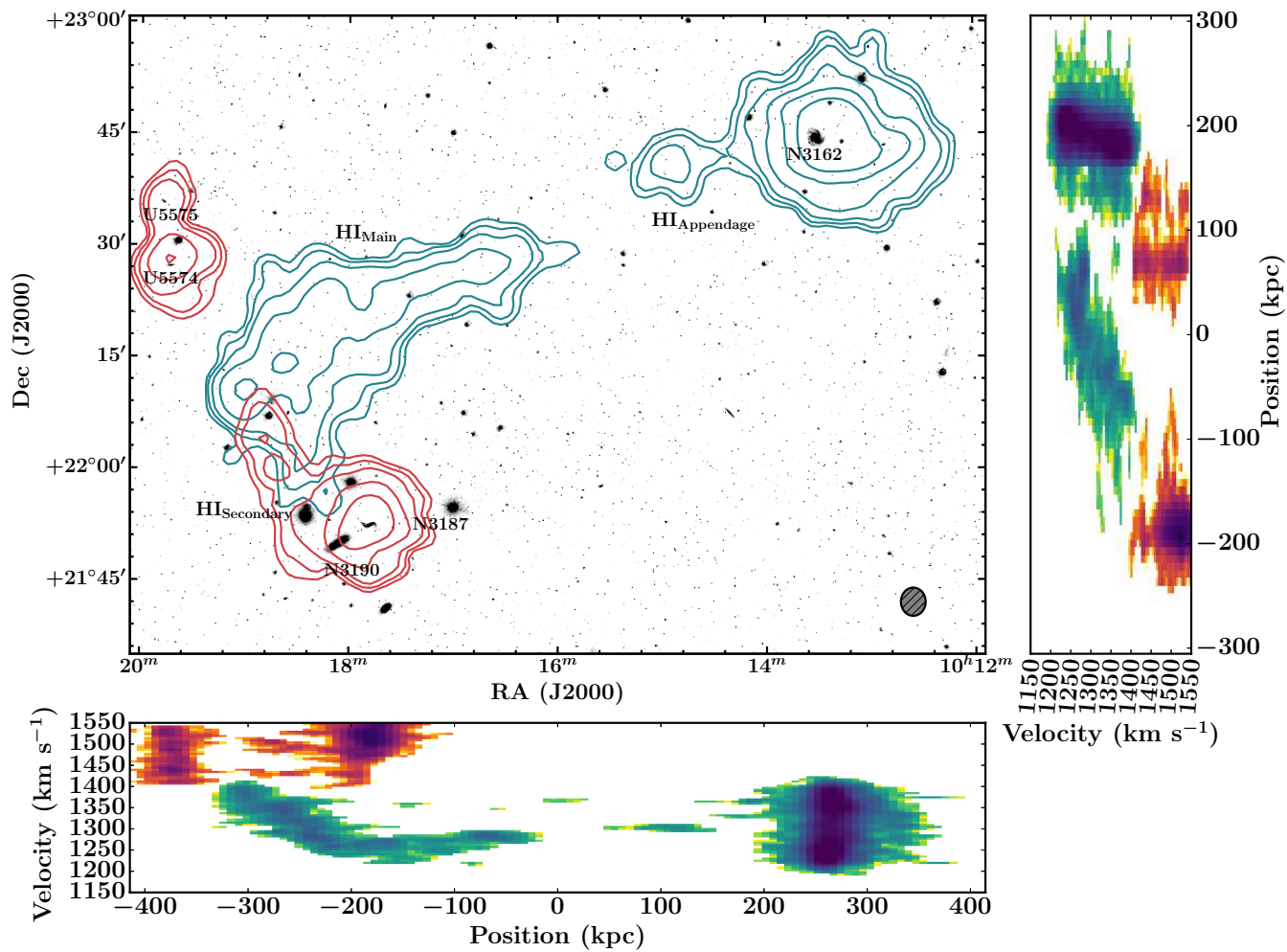} \\
\caption{ALFALFA moment 0 \hi\ contours, masked to highlight the primary \hi\ features in the  NGC~3190 group, overlaid on SDSS g-band optical images, with PV diagrams as in Figure \ref{hcg44map}.  The blue-green contours and blue to green coloured emission in the PV diagrams outline the entire main feature (HI$_{\rm Main}$), which is visible at velocities below 1400~\kms, and appears to be associated with  NGC~3162. The red contours and red to magenta coloured emission at velocities above 1400~\kms\ show the suggestion of a secondary feature (HI$_{\rm Secondary}$) at higher velocity, which may suggesting a connection between  NGC~3187 and UGC~5574, but this feature is only partially confirmed in deeper data (Hess et al., submitted). Contour levels are spaced logarithmically at 6, 12, 24, 48 96, and 192$\times10^{17}$ atoms~cm$^{-2}$ (assuming the \hi\ uniformly fills the ALFA beam).
}
\label{hcg44tails}
\end{figure*}

Indeed, since  HI$_{\rm Main}$ lies within the disk velocities of both NGC~3162 and NGC~3190, it is reasonable to assume that the feature is related to both sources.  However,
The connection between the giant \hi\ feature and the galaxies of  HCG~44 is less clear. The moment 0 contours in Figure \ref{hcg44map} show a low column density bridge between the brightest parts of the tail and  HCG~44. However, when examined in velocity space, the ALFALFA data hint that the extended \hi\ may be composed of at least two superimposed kinematic features. Figure \ref{hcg44tails} presents the region surrounding HI$_{\rm Main}$, masked to contain just the emission clearly associated with the extended \hi\ structures. The blue-green contours show the \hi\ emission associated with HI$_{\rm Main}$ and HI$_{\rm Appendage}$ demonstrating that together they extend nearly continuously from NGC~3162 toward a point $\sim$20\arcmin\ north of HCG~44,  bending through velocity space from $\sim$1300~\kms\ near NGC~3162 down to $\sim$1250~\kms, and then receding to $\sim$1400~\kms.  Near the SE end of HI$_{\rm Main}$ the \hi\ appears to reach south from the main feature toward NGC~3190, potentially suggesting a connection between HI$_{\rm Main}$ and the compact group.  However, combination of the ALFALFA data with the WSRT data from \cite{serra13a} and with recently obtained KAT-7 data (Hess et al., submitted) shows no strong evidence that HI$_{\rm Main}$ is connected to HCG~44 in the deeper cube.

The red contours in Figure \ref{hcg44tails} show \hi\ emission summed over velocities above 1400~\kms. These reveal a potential  second, low column density, low signal to noise structure, hereafter HI$_{\rm Secondary}$, visible at higher velocities ($\sim$1500~\kms). The structure extends east from  NGC~3187 to encompass the cloud dubbed C$_S$ in \cite{serra13a}, and then north toward UGC~5574, UGC~5575, and AGC~201174, at matching recessional velocities. This suggestion, however, is tentative, since the structure is detected at low significance (the significance of this structure in Figure \ref{hcg44tails} is potentially exaggerated by the superposition of signal from multiple low significance features).  The combined ALFALFA, WSRT, and KAT-7 data (Hess et al., submitted) also shows a compact \hi\ cloud at the location of the northern peak in the red contours in Figure \ref{hcg44tails}, but do not show clear emission connecting it to C$_{\rm S}$ or NGC~3187. Thus, much of the extent of this structure is likely a result of beam smearing and artefacts in the ALFALFA data.  

Figure \ref{hcg44map} also shows the apparent detection of an extended feature around 10$^h$15$^m$ +21$^{\circ}$20\arcmin. Analysis of the ALFALFA spectrum at this position shows significant baseline fluctuations, consistent with low level RFI not masked by our other algorithms. Similarly, Figure \ref{hcg44map} also shows the suggestion of a southern extension off NGC~3185, however, this extension is in the region of the cube with low weights due to significant RFI. Thus, further observations of these regions will be necessary to know if there is any real \hi\ emission to the south and southeast of HCG~44.

\subsection{The \hi\ mass budget in the  NGC~3190 and  NGC~3227 intragroup medium} 
\label{results.mass}

The extended features in both the NGC~3190 and NGC~3227 groups contain a significant fraction of the total \hi\ in the group.
Tables \ref{N3227table} and \ref{N3190table} list the fluxes and \hi\ masses for all sources detected by ALFALFA in the NGC~3227 and NGC~3190 groups respectively, assuming a common distance of 25~Mpc 
 (note that under this assumption, the NGC~3190 group contains 1.5$\times$ the \hi\ mass of  NGC~ 3227 group). 
Column 10 gives the full breakdown of \hi\ in each group, listing the percentage of \hi\ in each feature relative to the sum of each subgroup. 
The dark plumes in the  NGC~3227 group (HI$_{\rm North}$, HI$_{\rm South}$, and HI$_{\rm NE}$) have a combined log \hi\ mass of 9.47, 40\% of the total \hi\ in the group, and 53\% of the gas associated with NGC~3226/7. 
Only slightly less massive, HI$_{\rm Main}$ in the  NGC~3190 group has a log \hi\ mass of 9.05, making up 10\% of the total group \hi\ mass. However, 40\% of the \hi\ mass of the  NGC~3190 group is contained in  NGC~3162; HI$_{\rm Main}$ is nearly 30\% of the combined \hi\ mass of the galaxies in  HCG~44. 
For comparison, the Leo Ring makes up 24\% of the \mhi\ in the M96 group, and tidal plumes in the Leo Triplet constitute 14\% of the \hi\ mass in the M66 group \citep{stierwalt09a}.

Comparisons to synthesis imaging can give a sense of the amount of low surface density gas in these intragroup features, but direct comparison is difficult due to significant differences in the data and the subsequent analysis.
Tables \ref{N3227table} and \ref{N3190table} give synthesis measurements where available.  WSRT observations recover 94\% of the flux in HI$_{\rm NE}$, and 74\% of the flux in HI$_{\rm South}$ in the NGC~3227 group. However, the large extent of the dark \hi\ features necessitates large primary beam corrections for the portions of the sources near the edge of the beam, which means these results should not be over interpreted. For example, WSRT measures the flux of HI$_{\rm South}$ to be 5.1$\pm$0.2~Jy~\kms. However, if one only measures the flux outside the \cite{mundell95a} observation, and combines it with the \cite{mundell95a} VLA flux of 2.42~Jy~\kms, the summed synthesis observations give a flux of 6.0~Jy~\kms. It is also worth noting that \cite{mundell95a} measured a larger flux for HI$_{\rm North}$ than ALFALFA. However, the flux for HI$_{\rm North}$ is significantly blended with that of  NGC~3227 in the ALFALFA data, so the artificially low flux of HI$_{\rm North}$ is likely due to conservative deblending. 
A reasonable deduction from these data is that a significant portion of the flux in the  NGC~3227 subgroup is coming from the higher column density gas detected in synthesis observations.
 
ALFALFA detects fluxes significantly higher than those reported from WSRT in \cite{serra13a}, with the WSRT observations recovering between 43\% and 88\% of the ALFALFA flux for galaxies in  HCG~44. However, it is worth noting that the ALFALFA fluxes for  NGC~3187 and  NGC~3190 individually contain significant uncertainty, since their emission is blended in the ALFALFA cubes. It is also important to note that the flux of NGC~3185 is likely significantly affected by RFI, in spite of care during source parameter extraction to mitigate this effect. The range in recovered fluxes may suggest that  HCG~44 is surrounded by significant lower column density gas. 

Notably, ALFALFA detects the entirety of the \hi\ tail reported in \cite{serra13a}, and is unaffected by beam attenuation. \cite{serra13a} report a total tail log \hi\ mass of 8.71 after including HIPASS emission detected in re-reduced cubes. ALFALFA measures a total log \hi\ tail mass of 9.05$\pm$0.03, 2.2$\times$ higher than the WSRT+HIPASS measurement, significantly more than their stated typical flux uncertainties of 10-20\%.

\section{Discussion}
\label{discussion}

The ALFALFA maps of the NBG~21-6 region underscore the need for sensitive, wide field surveys to obtain a complete understanding of ongoing galaxy evolution in groups. Specifically, the detection of \hi\ tails and bridges on scales larger than the group virial radius suggests strong ties between the evolution of galaxies in this region, and their group environment. Here we discuss the regional context of the sources presented above as evidence of ongoing hierarchical galaxy assembly.

\subsection{ HCG~44 in the Context of the  NGC~3190 Group}
\label{discussion.compactgroups}

Previous studies of the \hi\ in the NGC~3190 group have focused on understanding the evolutionary state of HCG~44 in the context of of other compact groups. \cite{verdes-montenegro01a} propose an evolutionary sequence where a compact group's \hi\ deficiency \citep{haynes84a} increases with time due to multiple tidal interactions,  and find the the \hi\ deficiency of the members of HCG~44 to be high relative to other compact groups.\footnotemark[1] 
Similarly, \cite{borthakur10a} and \cite{borthakur15a} use the Green Bank Telescope to find evidence of diffuse \hi\ in compact groups, but find that most of the gas in a 180$\times$180~kpc (25\arcmin$\times$25\arcmin) region surrounding the central galaxies of HCG~44 originates in high column density, disk-like structures.
However, the apparent connection between NGC~3162 and HCG~44 in the ALFALFA data suggests that HCG~44 is best understood in the context of the entire NGC~3190 group. 

\footnotetext[1]{\cite{verdes-montenegro01a} measure the \hi\ deficiency of NGC~3185 and 3190 to be +1.0 (10\% of the expected gas), and the \hi\ deficiency of  NGC~3187 to be +0.4 (40\% of the expected gas). \cite{serra13a} get values a factor of 2 more deficient using the relations of \cite{toribio11b}, but we note that this difference can be explained by a typo swapping radius and diameter in the \cite{toribio11b} relations.}

Using \hi\ synthesis maps, \cite{serra13a} propose that the large \hi\ feature originated due to tidal stripping, and suggest two specific origins: 1) within the compact group due to an interaction between NGC~3190 and NGC~3187, or 2) from an interaction between NGC~3162 and the members of the compact group during a close flyby. In the context of the entire group, the ALFALFA data strongly favour the second hypothesis.

\cite{serra13a} suggest that if the \hi\ feature originated in HCG~44, it could at least partially explain the observed \hi\ deficiency. 
However, NGC~3162 dominates the \hi\ mass budget of the NGC~3190 group, with more than 150\% of the \hi\ mass contained in NGC~3185, NGC~3187, and NGC~3190 combined. While a highly disruptive event could remove $\sim$50\% of the gas from e.g., NGC~3187, tidal forces should more easily remove $\sim$20\% of the gas from the outskirts of an \hi-rich source like NGC~3162. Moreover, even after including the entire mass of HI$_{\rm Main}$, HCG~44 is still \hi\ deficient by more than a factor of six. This, suggests the members of HCG~44 were likely gas poor prior to an interaction with NGC~3162, which would make it even more difficult to remove the remaining, more tightly bound gas. Indeed, the potential detection of separate \hi\ features like HI$_{\rm Secondary}$ is consistent with the idea that the group has experienced other tidal interactions. Thus, the group context suggests a scenario where  the interaction with NGC~3162 is only one of the interactions responsible for the removal of gas from the members of HCG~44, and that NGC~3162 is the primary source of the gas in the \hi\ feature, potentially delivered on a first infall to the group. 

Further, the scale of the interactions surrounding HCG~44 suggests interesting questions for studies of other compact groups. Since not all tidal interactions favour the creation of massive tidal tails, the existence of a $\sim$600~kpc tail suggests that groups with less fortuitously aligned interactions on similar scales should exist. Studies like \cite{walker16a} have found only weak correlations between the total \hi\ in compact groups (as observed with GBT) and the properties of individual galaxies. However, the GBT has a 9\arcmin\ beam per pointing, and thus may miss significant group flux. In the case of HCG~44 (log $M_{\rm \footnotesize\textsc{H\,i},total}$ = 9.45), the entire compact group fits within just a couple of GBT beam pointings, but neither the large \hi\ tail (log \mhi\  = 9.05) nor  NGC~3162 (log \mhi\  = 9.65) would be detected. 

Thus, future wide-field surveys will be important for understanding the regions surrounding compact groups. While HCG~44 may be an extreme example, it presents a cautionary note that a full understanding of compact group scaling relations will likely require careful consideration of the surrounding $\sim$1~Mpc via deep, wide field \hi\ surveys.

\subsection{Interactions in their Group Context}
\label{discussion.galsingroups}

As discussed in section \ref{results.arp94}, the \hi\ features in the NGC~3227 group suggest potential associations between NGC~3227 and other group members, including NGC~3213, UGC~5653, and the dwarf galaxies in close spatial proximity to the plumes. Similarly, the apparent connection between NGC~3162 and NGC~3190 makes it highly probable that they are at the same distance. 
In regions of the local universe with limited distance information, upcoming \hi\ surveys will be an important tool for connecting sources with large distance uncertainties to those with primary distances. 
For example, a common way of determining group distances is to examine the Tully-Fisher relationship for the sources in the region (e.g. Figure \ref{TF} and the corresponding discussion in Section \ref{discussion.bigpicture}). However, stringent selection of galaxies for accurate distance determination (e.g. mostly edge on, similar Hubble types, etc.) often leaves too few sources for reliable distance determination, especially in compact groups where interactions may contribute to increased scatter in the relation. In these groups \hi\ mapping coupled with stellar distance estimates from JWST will be an important source of reliable distances. 

The potential tidal connection between sources separated by $\sim300$~kpc has potential consequences for searches for tidal dwarf galaxies. \cite{mundell04a} report a potential TDG candidate in close proximity to NGC~3226/7. While the plumes in the NGC~3227 group do not show clear evidence of TDGs further out at our current resolution, they do indicate that searches for tidal dwarf galaxies may require imaging at significant separations from their parent galaxies. For example, \cite{lelli15a} analyse VLA observations of 6 merging systems containing TDGs at cz$<$5000~\kms. The edges of the extended \hi\ emission in the NGC~3227 group would still extend just beyond the VLA primary beam if it were at  D=70~Mpc (5000~\kms). More, numerical simulations predict differing numbers of potential tidal substrates per system (e.g. \citealp{bournaud06a}; \citealp{yang14a}), so understanding the full scale of each interaction will be important in validating these predictions.

\subsection{Hierarchical Structure Formation in the NBG 21-6 Region}
\label{discussion.bigpicture}

The assembly of structures via mergers at all scales is a key prediction of hierarchical structure formation models. 
Thus, the discovery of two large tidal systems in relatively close proximity ($\sim$1~Mpc) motivates an examination of the surrounding region as a potential example of galaxy evolution on multiple scales.

\begin{figure}
\centering
\includegraphics[width=1.0\columnwidth]{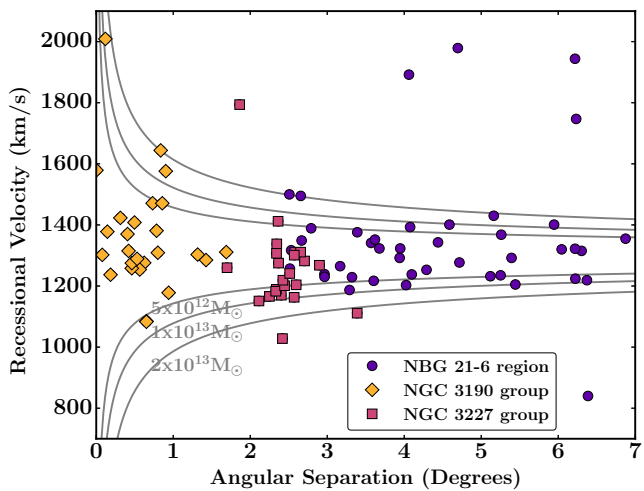}
\caption{Angular separation on the sky versus recessional velocity for galaxies in the NBG 21-6 region, showing the dynamics of the larger region. Separations are measured from the center of  HCG~44, which we take to be 154.5+22. Green squares are the galaxies associated with the NGC~3227 group, and red diamonds are those associated with  the NGC~3190 group. Blue circles are the other galaxies within 7 degrees of the center of  HCG~44.  At a distance of 25~Mpc, 7 degrees corresponds to $\sim$3~Mpc. Grey lines show simple caustic curves calculated using $v = \sqrt(GM/2r)$ (see Section \ref{discussion.bigpicture}).
}
\label{RVplot}
\end{figure}

As discussed in section \ref{GH58}, the grouping of galaxies in the NBG~21-6 region depends on the scale used in the grouping algorithm. However, the histogram in Figure \ref{regionhist} suggests a potential connection between galaxies in the broader region.
Figure \ref{RVplot} plots the recessional velocities from Figure \ref{regionhist} as a function of angular separation from the center of HCG~44 (HCG~44 was selected as the dynamical center since it has the largest velocity dispersion of the region). The narrow velocity structure ($\sigma\sim$115~\kms) over large angular scales suggests the region may be experiencing significant infall onto a larger, filamentary dark matter structure, though there are currently too few distance measurements to confirm this suggestion.

As a rough diagnostic tool, Figure \ref{RVplot} also shows simple caustic curves calculated using $v = \sqrt(GM/2r)$ and assuming all sources are at a distance of 25~Mpc. The distribution of sources in the region approximately match the velocity distribution expected for a sources moving in a potential of M = 2$\times10^{13}$\msun. Given the assumption of uniform distance and that this system is likely far out of equilibrium, these curves are at best suggestive. However, simple calculations of the dynamical mass of the  NGC~3190 and and NGC~3227 groups using the median mass calculation defined in \cite{heisler85a} estimate masses of 8.5 and 3.3$\times10^{12}$\msun\ respectively. We speculate that these estimates are consistent with a picture of the NGC~3190 and NGC~3227 groups as part of larger structure, falling toward the center of a larger common potential. Indeed, under the assumption that the two groups are moving toward each other, each with a velocity $v = \sqrt(G\times2\times10^{13}$\msun/($2\times1$~Mpc)) $\approx$ 200~\kms, the two systems will merge in $\sim$3-4~Gyr. 

\begin{figure}
\centering
\includegraphics[width=1.0\columnwidth]{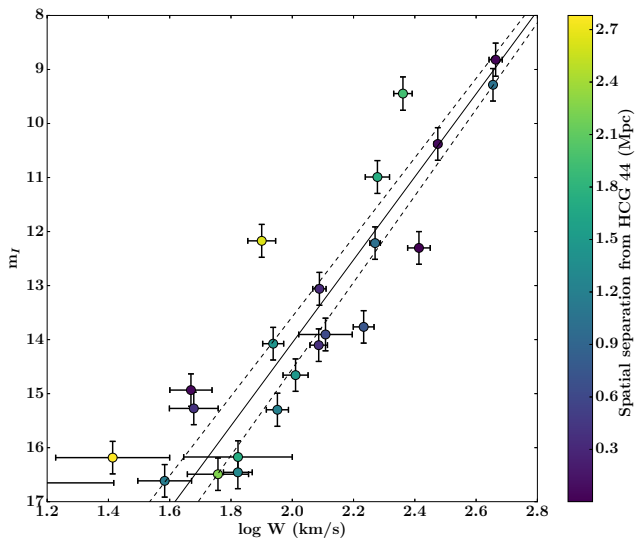}
\caption{Tully Fisher relation for the galaxies within the $2\times10^{13}$\msun\ caustic curves within R=3~Mpc of  HCG~44, demonstrating the difficulty of TF methods for determining group distance. The observed sources are consistent with all being at the same distance, but due to the lack of qualifying sources with current survey data, this result is simply suggestive.
Optical data were taken from the NASA-Sloan Atlas, and combined with the \hi\ data from ALFALFA. The sample was corrected and culled following the procedure outlined in \citet{giovanelli97a}, and discussed in section \ref{discussion.bigpicture}. The black line is the I-band relation from \citet{giovanelli97c} for a distance of 25~Mpc, with the dashed lines representing the 1$\sigma$TF error budget discussed in \citet{giovanelli97b}.  The color axis represent the distance of the source from the center of HCG~44.
}
\label{TF}
\end{figure}

If all the galaxies in the region are actually part of the same structure and at a similar distance, we would expect them to follow the Tully-Fisher relation between their apparent magnitude and their \hi\ velocity widths. Figure \ref{TF} shows the Tully-Fisher relation for sources contained within by the 2$\times10^{13}$\msun\ caustics in this 6~Mpc diameter region. Optical magnitudes were taken from the NASA-Sloan Atlas\footnotemark[1] where available, since the NSA compensates for significant shredding issues in the SDSS pipeline magnitudes. They were converted to I-band magnitudes using the Lupton (2005) relations from the SDSS website, and corrected for internal and galactic extinction following \cite{giovanelli97a}. NSA inclinations were checked by eyeball measurements, and any sources with (b/a)$_{NSA}$-(b/a)$_{eye}>0.2$ were removed. \hi\ velocity widths from ALFALFA were converted to total widths following \cite{giovanelli97a}, and all sources with i$<$40 were removed from the sample. Assuming all sources are the same distance and follow the Tully-Fisher relation from \cite{giovanelli97a}, we estimate the best fit distance to be 24.7$\pm$1.1~Mpc using a single parameter orthogonal distance regression fit to the data. 

\footnotetext[1]{http://www.nsatlas.org/}
The black line shows the I-band Tully-Fisher relation and error budget from \cite{giovanelli97a} and \cite{giovanelli97b}, for a distance of 25~Mpc. The scatter around the relation is consistent with all sources at the same distance for the small number of sources available. 
However, while this result is consistent with the interpretation of (almost) all of the sources in the region being at the same distance, the lack of sources makes it impossible to strongly differentiate between that and significant distance scatter,  since a source with W = 160~\kms\ in the background by 5~Mpc, would only lie $\sim$1$\sigma$ off the relation. Thus, this calculation demonstrates the difficulty of using secondary relations to determine group distance. 
Significant source statistics to perform this kind of analysis will have to await next generation deep \hi\ and optical surveys, and improved understanding of the faint end of the baryonic Tully Fisher relation. 


The collective optical properties of galaxies in the NBG~21-6 region are consistent with expectations for galaxies experiencing significant large scale infall in a hierarchical formation model. Most of the sources in the NBG~21-6 region are dwarfs with a r-band absolute magnitude of M$_r$$>$-18. Most sources in the region are \hi\ rich (ALFALFA detects 73\% of the galaxies with measured redshifts in NBG~21-6), and significantly bluer than the average SDSS population. Moreover, sources within the NGC~3190 and NGC~3227 groups may be slightly more processed than the other galaxies in the region. ALFALFA detects 81\% of the 43 galaxies outside the two group, compared with 63\% of the 36 galaxies inside the groups, while examination of the colour distribution hints of that galaxies in the two groups appear to be somewhat redder than the other sources in the region (though this result is only marginally significant: a KS test on the two colour distributions gives $p=0.08$). The largest galaxies near the centers of the NGC~3190 and NGC~3227 groups appear to be the reddest sources in the region, and indeed, \cite{appleton14a} discuss the importance of the green valley colour of  NGC~3226, demonstrating that the colour cannot be explained by a recent resurgence of star formation, but rather must result from quenching of fairly recent star formation.
 
These observations together paint an observational picture that mirrors recent simulations of group and cluster formation in a hierarchical context. 
HCG~44, which shows the largest velocity spread in the region (NGC~3185 and NGC~3187 are separated by $\sim$360~\kms ), sits at the bottom of a larger potential well, with the other members of  NGC~3190 delivering gas as they fall in. The merging pair NGC~3226/3227 rests near the bottom of a smaller potential that is falling toward the  NGC~ 3190 well. Currently out of equilibrium, the region will continue to group and virialize as it makes its journey toward the Virgo cluster. 

The lack of primary distance information and the poorly populated nature of these groups cautions against over-interpretation of the current data, since without accurate distance measurements, the possibility remains that velocity crowding gives the appearance of connection between unrelated sources. Further, in a hierarchical model, it is possible that interactions play a significant role in increasing the velocity dispersion of  the NGC~3190 and NGC~3227 groups, effecting mass estimates.  However, this caution suggests important synergies between upcoming optical surveys (which can deliver more and better understood primary distance measurements), and upcoming sensitive, wide field \hi\ surveys in understanding this fortuitously aligned group, and others like it. 

\section{Conclusions}
\label{conclusions}

In this paper we presented sensitive, wide-field \hi\ imaging of the NBG 21-6 region of the Leo Cloud of galaxies, observed as part of the ALFALFA \hi\ survey, and follow up imaging of several regions using WSRT.  We detect intra-group plumes that each extend over $\sim$2 degrees  ($\sim$600~kpc), far beyond the primary beams of current synthesis telescopes. These features reveal interactions on larger scales than initially anticipated, providing an important tool for future modeling the recent history of these systems. 

Specifically, the main conclusions of this paper are:
\begin{enumerate}
\item The detection of multiple $>$300~kpc \hi\ appendages in a 2x2~Mpc area of the sky mapped by ALFALFA.  We find that the large gas tail to the north of  HCG~44 detected in \cite{serra13a} is likely associated with NGC~3162, as speculated by those authors, and find a tentative detection of a second, superimposed tail associated with NGC~3187 and UGC~5574.  We additionally find the \hi\ plumes in the NGC~3227 group presented in \cite{mundell95a} extend far beyond the observed primary beam, and report the detection of a third, clumpier feature to the NE of the system, which displays a prominent kink in WSRT observations, suggestive of additional interactions.
\item \hi\ features without associated stellar counterparts make up a significant component of the group gas content in the groups considered here. The intra-group gas makes up 10\% of the  NGC~3190 group and 40\% of the NGC~3227 group. 
\item The NGC~3190 group (containing HCG~44) and the NGC~3227 group may be part of a larger, dynamically young region experiencing active infall, and may merge in $\sim$3-4~Gyr. Examination of the broader group dynamics and optical colors suggests that its relatively local proximity and projection on the sky make the NBG 21-6 region an important laboratory for studying the role of ongoing galaxy interactions in the progression of group galaxies from the blue cloud to the red sequence. 
\end{enumerate}
Our observations emphasize the importance of wide-field, high sensitivity \hi\ mapping. These massive, extended \hi\ features were found in a region that had been extensively studied in \hi; without the powerful combination of sensitivity, resolution, and field of view of ALFALFA we would not have been able to trace galaxy interactions in these large scales. Wide field, deep mapping with next generation \hi\ detectors like the Square Kilometer Array Pathfinders (see \citealp{giovanelli16a}) will be necessary for a full understanding of galactic evolution in the group environment. 

\section*{Acknowledgements}
The authors acknowledge the work of the entire ALFALFA
collaboration in observing, flagging, and extracting sources in this field. 
The ALFALFA team at Cornell is supported by NSF grants AST-0607007 
and AST-1107390 to RG and MPH and by grants
from the Brinson Foundation. EAKA is supported by TOP1EW.14.105, which is financed by the Netherlands Organisation for
Scientific Research (NWO). The authors would like to thank the referee for useful comments that helped improve the quality of the paper. LL would like to thank Michael G. Jones and Gregory Hallenbeck for several useful discussions.

This research used data from the Sloan Digital Sky Survey, funded by
the Alfred P. Sloan Foundation, the participating institutions, the
National Science Foundation, the U.S. Department of Energy, the
National Aeronautics and Space Administration, the Japanese
Monbukagakusho, the Max Planck Society, and the Higher Education
Funding Council for England.

The Westerbork Synthesis Radio Telescope is operated by the ASTRON (Netherlands Institute for Radio Astronomy) with support from the Netherlands Foundation for Scientific Research (NWO).



\bibliography{mybib}
\bibliographystyle{mnras}


\begin{table*}
\caption{Galaxies in the NGC~3227 Group. Column 1: ID Number from the Arecibo General Catalog; column 2: alternate name; column 3: RA and Dec correspond to the optical center of each galaxy, except for \hi\ only features, where the coordinates give the position of the peak flux; column 4: heliocentric optical recessional velocity. For all sources detected in \hi, the value is from the ALFALFA catalog,
else the value is from the Arecibo General Catalog; column 5: \hi\ line width from ALFALFA, measured at the 50\% flux level; note that AGC~5620 and HI$_{\rm North}$ are blended, so the fluxes contain addition systematic uncertainty; column 6: $\int{SdV}$ as measured in the ALFALFA data;
column 7: $\int{SdV}$ as measured in synthesis data, corrected for primary beam attenuation;
column 8: Source of the synthesis measurement. Sources labelled L16 are from WSRT measurements presented in this paper, and 
those labeled M95 are VLA measurements from \citet{mundell95a};
column 9: \hi\ Masses assume a distance of 25~Mpc for all sources;
column 10: Fraction of group \hi, $F_{\rm \textsc{H\,i},galaxy}$/$F_{\rm \textsc{H\,i},group}$, in percent.
$^*$Classified as a member of the NGC~3190 group by \citet{makarov11a}
$^{\dagger}$Classified as a background source by \citet{makarov11a}
Note: Sources below the horizontal line are not detected at the ALFALFA sensitivity limit, which is a function of W$_{50}$ and the rms at the position 
of the source (see eq. 2 in \citet{haynes11a}). For a 50~\kms\ wide source, the average 4$\sigma$ ALFALFA flux upper limit 
is 0.3 Jy-km/s, which translates into a fractional limit of $<$0.6\% of the total group mass for each undetected source. }  
\label{N3227table}
\begin{tabular}{lcccrrcccr}
\hline
AGC  & Name & Coordinates & c$z$ &W$_{\rm 50}$ & $F_{\rm \textsc{H\,i},ALFA}$ & $F_{\rm \textsc{H\,i},SYN}$  & Source & log \mhi & $f_{\rm \textsc{H\,i}}$\\
  &       & J2000 & \kms & \kms & Jy-\kms & Jy-\kms &   & \msun  & \% \\
(1) & (2) & (3) & (4) & (5) & (6) & (7) & (8) & (9) & (10) \\
\hline
5590$^*$ & NGC 3213 & 10:21:17.4+19:39:04 & 1347$\pm$1 & 134$\pm$2 & 1.62$\pm$0.06 & &  & 8.38 & 3.2\\
5620 & NGC 3227 & 10:23:30.5+19:51:53 & 1126$\pm$2 & 412$\pm$3 & 17.52$\pm$0.10 & 14.5& M95 & 9.41 & 35.0\\
5629$^*$ &  & 10:24:12.9+21:03:01 & 1238$\pm$1 & 113$\pm$2 & 3.58$\pm$0.06 & &  & 8.72 & 7.2\\
5653 & I610 & 10:26:28.3+20:13:42 & 1168$\pm$2 & 296$\pm$4 & 1.64$\pm$0.06 & &  & 8.38 & 3.3\\
5675 & M+327047 & 10:28:30.0+19:33:46 & 1106$\pm$1 & 83$\pm$2 & 4.55$\pm$0.06 & &  & 8.83 & 9.1\\
202045$^{\dagger}$ & D568-04 & 10:20:56.1+20:09:21 & 1557$\pm$2 & 73$\pm$4 & 1.31$\pm$0.04 & &  & 8.29 & 2.6\\
HI$_{\rm NE}$ &  & 10:25:37.4+20:20:34 & 1210$\pm$2 & 57$\pm$4 & 10.18$\pm$0.21 & 9.61$\pm$0.39 & L16 & 9.18 & 20.3\\
HI$_{\rm North}$ &  & 10:23:25.7+20:00:01 & 1305$\pm$33 & 134$\pm$16 & 2.88$\pm$0.16 & 3.3& M95 & 8.63 & 5.8\\
HI$_{\rm South}$ &  & 10:22:50.5+19:27:48 & 1292$\pm$5 & 49$\pm$5 & 6.77$\pm$0.21 & 5.12$\pm$0.2 & L16 & 9.00 & 13.5\\
\hline
\multicolumn{8}{l}{NGC 3227 group total:} & 9.87 & 1.0 \\
\hline
5617 & NGC 3226 & 10:23:27.0+19:53:54 & 1275 & ... &  ... & &  &  ...  &  ... \\
718673 &  & 10:23:15.4+20:10:40 & 1151& ... &  ... & &  &  ...  &  ... \\
718681 &  & 10:23:21.6+20:01:38 & 1166& ... &  ... & &  &  ...  &  ... \\
718719 &  & 10:24:34.7+20:01:57 & 1028& ... &  ... & &  &  ...  &  ... \\
718778 &  & 10:26:31.0+20:16:60 & 1163& ... &  ... & &  &  ...  &  ... \\
739353 &  & 10:23:22.4+19:54:51 & 1338& ... &  ... & &  &  ...  &  ... \\
739467 &  & 10:27:37.5+20:04:43 & 1268& ... &  ... & &  &  ...  &  ... \\
\hline 
\end{tabular}
\end{table*}


\begin{table*}
\caption{Galaxies in the NGC~3190 Group. Column Definitions are the same as Table 1. Note that, as in Table 1, distance dependent parameters are calculated using a distance of 25~Mpc for all sources. Measurements labeled S13 are WSRT measurements
from \citet{serra13a}. Note that AGC~5556 and 5559 are blended in the ALFALFA data. For a 50~\kms wide source, the average 4$\sigma$ ALFALFA flux upper limit of 0.3 Jy-km/s, 
translates into a fractional limit of $<$0.4\% of the total group mass for each undetected source. 
$^*$Classified as a member of the NGC~3227 group by \citet{makarov11a} 
$^{\dagger}$Classified as a background source by \citet{makarov11a}. } 
\label{N3190table}
\begin{tabular}{lcccrrcccr}
\hline
AGC  & Name & Coordinates & c$z$ &W$_{\rm 50}$ & $F_{\rm \textsc{H\,i},ALFA}$ & $F_{\rm \textsc{H\,i},SYN}$  & Source & log \mhi & $f_{\rm \textsc{H\,i}}$\\
  &       & J2000 & \kms & \kms & Jy-\kms & Jy-\kms &   & \msun  & \% \\
(1) & (2) & (3) & (4) & (5) & (6) & (7) & (8) & (9) & (10) \\
\hline
5510 & NGC 3162 & 10:13:31.6+22:44:13 & 1302$\pm$1 & 177$\pm$2 & 30.48$\pm$0.08 & &  & 9.65 & 40.0\\
5524 & 123-027 & 10:14:21.8+22:07:28 & 1636$\pm$1 & 180$\pm$2 & 2.73$\pm$0.07 & &  & 8.60 & 3.6\\
5544 & NGC 3177 & 10:16:34.0+21:07:23 & 1310$\pm$2 & 181$\pm$4 & 3.76$\pm$0.07 & 2.11& S13 & 8.74 & 4.9\\
5554 & NGC 3185 & 10:17:38.6+21:41:16 & 1230$\pm$1 & 253$\pm$2 & 3.58$\pm$0.08 & &  & 8.72 & 4.7\\
5556 & NGC 3187 & 10:17:47.9+21:52:23 & 1586$\pm$2 & 219$\pm$4 & 11.19$\pm$0.09 & 8.22& S13 & 9.22 & 14.7\\
5559 & N3189/90\footnotemark[1] & 10:18:05.5+21:49:51 & 1310$\pm$4 & 463$\pm$7 & 4.53$\pm$0.12 & 4.0& S13 & 8.82 & 5.9\\
5574 & M+425001 & 10:19:43.0+22:27:06 & 1462$\pm$5 & 130$\pm$10 & 2.30$\pm$0.08 & &  & 8.53 & 3.0\\
5575 &  & 10:19:46.9+22:35:39 & 1471$\pm$3 & 123$\pm$5 & 1.28$\pm$0.07 & 0.84& S13 & 8.28 & 1.7\\
200162 & 123-024 & 10:12:52.5+22:43:21 & 1285$\pm$11 & 128$\pm$23 & 1.48$\pm$0.07 & &  & 8.34 & 1.9\\
200255$^*$ & 124-001 & 10:19:01.5+21:17:01 & 1083$\pm$3 & 50$\pm$6 & 2.65$\pm$0.06 & &  & 8.59 & 3.5\\
201052$^{\dagger}$ &  & 10:17:23.3+21:47:58 & 1943$\pm$1 & 42$\pm$3 & 2.38$\pm$0.06 & 1.03& S13 & 8.55 & 3.1\\
201174 & F500-4 & 10:19:19.2+22:42:05 & 1573$\pm$3 & 51$\pm$6 & 0.57$\pm$0.06 & &  & 7.92 & 0.7\\
201337 & 123-035 & 10:17:39.5+22:48:35 & 1178$\pm$3 & 183$\pm$7 & 1.20$\pm$0.08 & &  & 8.25 & 1.6\\
202134 & Wa5 & 10:10:32.8+22:00:40 & 1280$\pm$12 & 79$\pm$25 & 0.37$\pm$0.06 & &  & 7.74 & 0.5\\
HI$_{\rm Main}$ & T$_{\rm N}$ & 10:18:05.4+22:31:27 & 1302$\pm$5 & 115$\pm$3 & 7.66$\pm$0.43 & 2.79& S13 & 9.05 & 10.1\\
\hline
\multicolumn{8}{l}{NGC~3190 group total:} & 10.05 & 1.0 \\
\hline
5562 & NGC 3193 & 10:18:24.9+21:53:36 & 1378 &...&  ... & &  &  ...  &  ... \\
208748 & 124-001b & 10:19:00.8+21:16:55 & 1085&...&  ... & &  &  ...  &  ... \\
208761 &  & 10:19:28.6+21:11:24 & 1381&...&  ... & &  &  ...  &  ... \\
718513 &  & 10:17:17.6+22:09:39 & 1423&...&  ... & &  &  ...  &  ... \\
718530 &  & 10:18:22.8+21:23:31 & 1408&...&  ... & &  &  ...  &  ... \\
718531 &  & 10:18:40.9+21:22:49 & 1289&...&  ... & &  &  ...  &  ... \\
\hline
\end{tabular}
\end{table*}
\footnotetext[1]{NGC~3190 is referred to as NGC~3189 in some papers. According to \citet{serra13a}, NGC3189 is the southeast component of NGC~3190.}

\bsp	
\label{lastpage}
\end{document}